\documentclass[aps,twocolumn,prb,amsmath,amssymb]{revtex4-1}

\usepackage{amsmath}
\usepackage{amssymb}
\usepackage[varg]{txfonts}
\usepackage{color}
\usepackage[colorlinks]{hyperref}
\usepackage{tikz}
\usetikzlibrary{shapes}
\usepackage{xcolor}

\usepackage{graphicx}
\usepackage{bm}

\newcommand{\hexagon}{\mathord{\raisebox{0.6pt}{\tikz{\node[draw,scale=.65,regular polygon, regular polygon sides=6,fill=white](){};}}}}

\newcommand{\RNum}[1]{\uppercase\expandafter{\romannumeral #1\relax}}

\def\XXint#1#2#3{{\setbox0=\hbox{$#1{#2#3}{\int}$}
    \vcenter{\hbox{$#2#3$}}\kern-.5\wd0}}

\def\be{\begin{equation}}
\def\ee{\end{equation}}

\def\bi{\begin{itemize}}
    \def\ei{\end{itemize}}
\def\bn{\begin{enumerate}}
    \def\en{\end{enumerate}}
\def\bea{\begin{eqnarray}}
\def\eea{\end{eqnarray}}
\newcommand{\bpm}{\begin{pmatrix}}
    \newcommand{\epm}{\end{pmatrix}}

\def\ba{\begin{array}}
    \def\ea{\end{array}}
\def\bd{\begin{displaymath}}
\def\ed{\end{displaymath}}

\renewcommand{\imath}{\hspace{1pt}\mathrm{i}\hspace{1pt}}

\renewcommand{\vec}{\mathbf}

\begin{document}

\title{Designing $\mathbb{Z}_2$ and $\mathbb{Z}_2 \times \mathbb{Z}_2$ topological orders in networks of Majorana bound states }

\author{Fatemeh Mohammadi}
\affiliation{Department of Physics, Sharif University of Technology, Tehran 14588-89694, Iran}

\author{Mehdi Kargarian}
\email{kargarian@physics.sharif.edu}
\affiliation{Department of Physics, Sharif University of Technology, Tehran 14588-89694, Iran}

\begin{abstract}
Topological orders have been intrinsically identified in a class of systems such as fractional quantum Hall states and spin liquids. Accessing such states often requires extreme conditions such as low temperatures, high magnetic fields, pure samples, etc. Another approach would be to engineer the topological orders in systems with more accessible ingredients. In this work, we present networks of Majorana bound states, which are currently accessible in semiconductor nanowires proximitized to conventional superconductors, and show that the effective low-energy theory is topologically ordered. We first demonstrate the main principles in a lattice made of Kitaev superconducting chains comprising both spin species. The lattice is coupled to free magnetic moments through the Kondo interaction. We then show that at the weak coupling limit, effective ring spin interactions are induced between magnetic moments with a topological order enjoying a local $\mathbb{Z}_2 \times \mathbb{Z}_2$ gauge symmetry. We then show that the same topological order and also the $\mathbb{Z}_2$ one can be engineered in architecture patterns of semiconductor  nanowires hosting Majorana bound states. The basic blocks of patterns are the time-reversal Majorana Cooper boxes coupled to each other by metallic leads, and the Majorana states are allowed to tunnel to quantum dots sitting on the vertices of the lattices. In the limit of strong onsite Coulomb interactions, where the charge fluctuations are suppressed, the magnetic moments of dots on the square and honeycomb lattices are described by topologically ordered spin models with underlying $\mathbb{Z}_2$ and $\mathbb{Z}_2 \times \mathbb{Z}_2$ gauge symmetries, respectively. Finally, we show that the latter topological order can also be realized in a network of purely Majorana zero modes in the absence of coupling to quantum dots.                         

\end{abstract}

\maketitle

\section{Introduction}\label{Introduction}
Topological order is characterized by long-range patterns of entanglement, multiple degenerate ground states, and exotic quasiparticle excitations. The topologically protected ground states and nontrivial braiding statistics of quasiparticles provide a unique platform for fault-tolerant quantum computations\cite{DasSarma:RMP2008}. Notwithstanding massive efforts put forward in the last few decades, the experimental realization of quantum states with topological orders in a controlled way and accessible conditions has remained elusive. The realization of such states in fractional quantum Hall states at extreme conditions like the cryogenic temperatures and strong magnetic fields \cite{FQHE:1999}, or in the spin liquid material RuCl$_3$ in a narrow window of an in-plane magnetic field \cite{Matsuda:Nat2018}, obfuscates the practical use of these intrinsic topological states in designing quantum memories and gates. 

Alternatively, we may think of extrinsic models to simulate the exotic phases by combining a set of naturally accessible quantum systems. Recently, an idea based on the topological bulk proximity effect has been introduced by Heish, et al., \cite{Hsieh2016, Hsieh2017}. It is shown that a square lattice of Majorana fermions originating from a set of one-dimensional topological superconductors, when Kondo coupled to an otherwise lattice of free magnetic ions, induces abelian $\mathbb{Z}_2$ topological order in the magnetic system in the weak coupling limit. In another work by Terhal, et al., \cite{Terhal:PRL2012} it is shown that the low-energy model describing a network of hybridized Majorana fermions resembles that of toric code model \cite{KITAEV20032}. Also, a duality between interacting Majorana systems and some quantum models such as Ising gauge theories, transverse-field Ising model, and a spin compass model on the honeycomb lattice has been discussed in a work by Nussinov, et al., \cite{Nussinov:prb2018}.

In this work, we design several systems comprised of Majorana zero states to simulate another class of topologically ordered states with underlying $\mathbb{Z}_2 \times \mathbb{Z}_2$ gauge symmetry, the so-called color code model, with enhanced computational capabilities\cite{Bombin2006}. On a closed manifold with genus $g$, the model encodes $4g$ qubits, and when embedded on lattices with boundaries, a set of Clifford gates in quantum teleportation can be performed transversally, which is not possible with toric code model. The color code Hamiltonian describes the gapped low-energy excitations of a spin model with two-body interactions $\sigma^{\alpha}\sigma^{\alpha}$ ($\alpha=x,y,z$) on the ruby lattice \cite{Bombin2009, Kargarian_2010}, much like the toric code model as a description of the Kitave honeycomb model\cite{KITAEV20062}. While it might be challenging to find proper magnetic materials with underlying ruby lattice interactions calling for further materials investigations, here we pursue a different path to engineer such enriched topological order using networks of Majorana bound states.   

Let us give a brief synopsis of Majorana models developed in this work. (i) The first model consists of the Kitaev $p$-wave superconducting chains of up and down spins. The lattice sites are the crossing points of chains and are Kondo coupled to a honeycomb lattice of magnetic ions, each carrying spin-$1/2$; (ii) We then introduce a more feasible way of simulating the above model. Instead of the Kitaev chains, we demonstrate that the Majorana bound states appearing at the ends of semiconductor nanowires proximitized to conventional superconductors with time-reversal symmetry\cite{Qi2009, Zhang2013} can also induce topological orders on the underlying lattice of quantum dots with strong onsite Hubbard interaction. We first show that the formerly induced $\mathbb{Z}_2$ toric code model in Ref.[\onlinecite{Hsieh2017}] arises in a more accessible system, and then present an architecture of Majorana zero states to construct $\mathbb{Z}_2 \times \mathbb{Z}_2$ topological order; (iii) Topologically ordered states can also be created in purely Majorana models known as Majorana surface codes in a network of Majorana Cooper boxes (MCBs) \cite{Beri2012,Altland2013, Beri2013, Zazunov2014, Altland2014}. Each box consists of two Rashba semiconductor wires in proximity to an $s$-wave superconductor \cite{Lutchyn2010, Alicea2012}. In the presence of a Zeeman field, Majorana fermions appear at the end of wires, so the ground state of each box is four-fold degenerate with each pair belonging to a parity sector. By applying charging energy, a pair with definite parity is singled out, hence yielding an effective spin-half description for each box \cite{Fu2010, Zazunov2011, Plugge2016, Plugge2016_2, Plugge2016_3,thomson2018, Tsvelik2020}. To construct a Majorana fermion code with $\mathbb{Z}_2 \times \mathbb{Z}_2$ gauge symmetry, we first introduce a new MCB with four nanowires and then show that an arrangement of such islands on the vertices of a honeycomb lattice and tunnel couple them by metallic leads generate a Majorana fermion code\cite{Bravyi2010} with $\mathbb{Z}_2 \times \mathbb{Z}_2$ topological order. 

The paper is organized as follows. In Sec.\ref{superconductor_model}, we introduce a lattice of Kitaev chains coupled to magnetic moments. In Sec.\ref{TRMCB_model},  architecture patterns of Majorana bound states on square and honeycomb lattices, and their tunneling to quantum dots are presented, and in Sec.\ref{constructing_Z2timesZ2_model} realization of topological order in a lattice of Majorana states is investigated. Sec.\ref{conclusions} summarizes and concludes.    

\section{ $\mathbb{Z}_2 \times \mathbb{Z}_2$ topological order induced by topological superconductors}\label{superconductor_model}

\begin{figure}[t]
\center
\includegraphics[width=\linewidth]{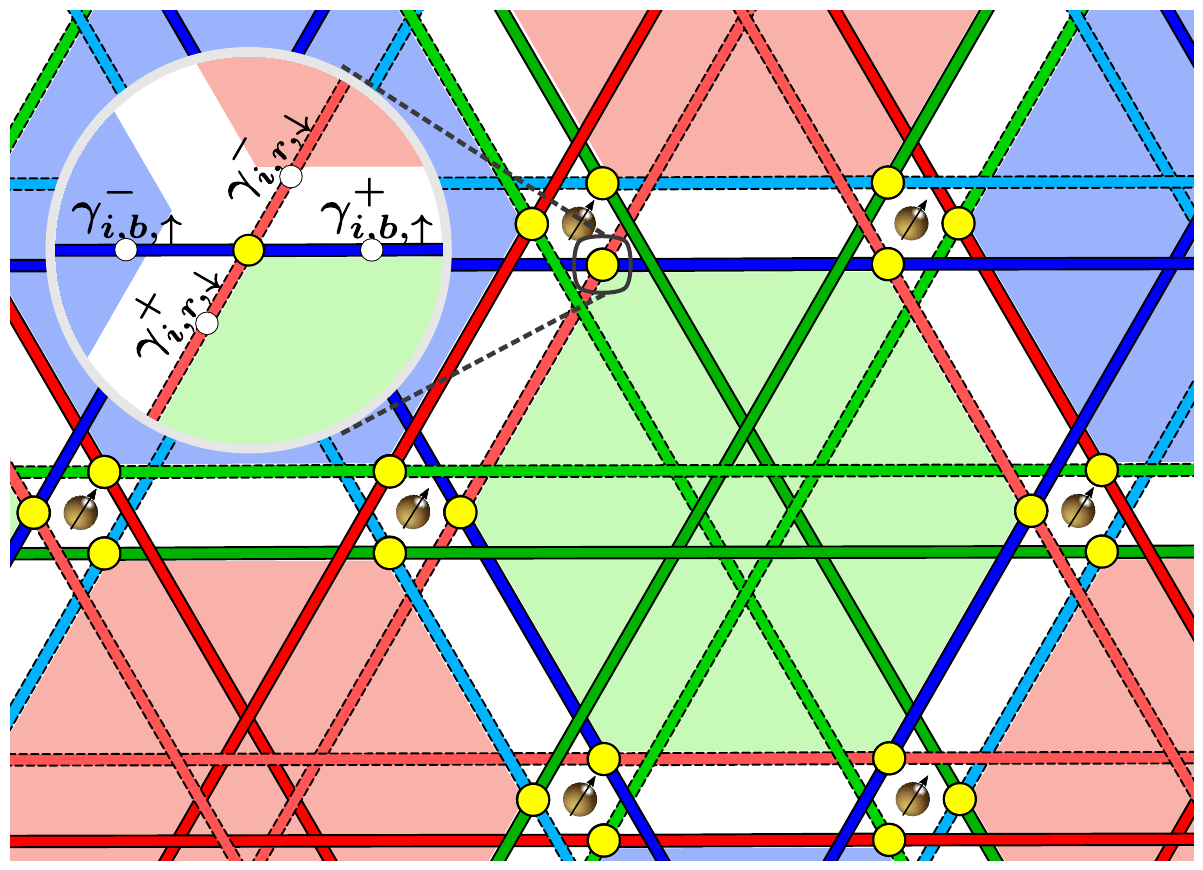}
\caption{Chains of the Kitaev topological superconductors. Dark (light) chains with solid (dashed) borders present 1D superconductors with spin up (down). Starting from horizontal red, green, and blue chains, other chains are obtained by rotating them by 60-degree. We color the hexagonal plaquettes by the color of chains crossed. Chains with different colors and spins cross at yellow dots making the sites of the lattice; this is system A. Brown spheres indicate magnetic moments coupled to lattice sites via Kondo interaction forming system B. This model induces a $\mathbb{Z}_2 \times \mathbb{Z}_2$ topological order in system B. The circle zoomed out reveals the representation of electrons in terms of Majorana fermions. \label{superconductor_scheme}}
\end{figure}

The first lattice model is shown in Fig.~\ref{superconductor_scheme}. The solid yellow vertices are the crossing points of Kitaev $p$-wave superconducting chains; colored solid chains carry only electrons with spin up while dashed chains carry only electrons with spin down. Hence each chain is effectively a spinless Kitaev chain which is topological in a suitable range of parameters. Chains with up and down spins appear in three colors, red, green, and blue, and are oriented 60-degree with respect to each other. Note that we just use colors to keep track of chains and their orientations. Only chains carrying up and down spins  with different colors are allowed to intersect and form lattice sites. We present them symbolically as follows:             

\begin{equation}\label{mapping_intersection}
\begin{split}
&\mathrm{TS}_{r, \uparrow}  \times  \mathrm{TS}_{g, \downarrow}\\
&\mathrm{TS}_{g, \uparrow}   \times \mathrm{TS}_{b, \downarrow}\\
&\mathrm{TS}_{b, \uparrow}   \times \mathrm{TS}_{r, \downarrow},
\end{split}
\end{equation}
where $\mathrm{TS}_{c,\sigma}$ indicate a topological superconductor chain with color $c = r,g,b$, and spin $\sigma = \uparrow, \downarrow$. $r$, $g$, and $b$ stand for red, green, and blue chains, respectively, and $\times$ shows the intersection between the corresponding chains. The lattice naturally admits to color the hexagonal plaquettes crossed by the chains with the same color.  

Given a colored chain indexed by the pair $\{c,\sigma\}$, the superconducting Hamiltonian reads as 
\begin{equation}\label{Kitaev_SC}
H_{c,\sigma}=\sum_{i} \left(-ta^{\dagger}_{i,c,\sigma}a_{i+1,c,\sigma}+\Delta a_{i,c,\sigma}a_{i+1,c,\sigma}+\mathrm{h.c.}  \right),
\end{equation}  
where $a^{\dagger} (a)$ denotes the fermionic creation (annihilation) operator. $t$ and $\Delta$ are hopping and superconducting pairing amplitudes, and for the sake of simplicity, we set $t=\Delta=1$ throughout, hence yielding a topological ground state for \eqref{Kitaev_SC}. The full Hamiltonian for the lattice in Fig.~\ref{superconductor_scheme} is then $H_{A}=\sum_{c,\sigma}H_{c,\sigma}$. Writting the complex fermions in terms of Majorana operators $a_{i, c, \sigma}=(\gamma_{i, c, \sigma} ^- + i \gamma_{i, c, \sigma}^+)/2$ satisfying  the Clifford algebra $\{\gamma_{i, c, \sigma}^s, \gamma_{i, c, \sigma}^{s'} \}=2\delta_{ss'}$ with $s,s'=\pm$ (see a lattice site zoomed out in  Fig.~\ref{superconductor_scheme}), the parent Hamiltonian $H_{A}$ is recast as        

\begin{equation}\label{A_superconductor}
\begin{split}
H_A = \sum_{i, c, \sigma} i \gamma_{i, c, \sigma}^+ \gamma_{i + 1, c, \sigma}^-.
\end{split}
\end{equation}

Here the sum over ``$i$" runs over sites of a colored chain. Note that this model is exactly solvable thanks to the bilinear structure of terms in $H_A$. Indeed, each term is a stabilizer operator consisting of two Majorana modes sitting on the link connecting neighboring sites: they commute with each other mutually and squares to identity individually. The excitations are gapped, and the ground state of this Hamiltonian is

\begin{align}\label{A_superconductor}
|\psi_{A}\rangle = \prod_{i, c, \sigma} \left(1 - i\gamma_{i, c, \sigma}^+ \gamma_{i + 1,c,\sigma}^-\right)|0\rangle,
\end{align}
with ground state energy $E_{gs}=-6N$ where $N$ is the number of lattice sites.
To construct an effective spin model with topological order, we consider a lattice of magnetic ions with spin-$1/2$ residing on the vertices of a honeycomb lattice, the brown dots shown in Fig.~\ref{superconductor_scheme}. We call it system $B$ with a trivial Hamiltonian $H_{B}=0$. Each magnetic ion is Kondo coupled to the superconducting sites as
 
\begin{equation}\label{kondo_coupling}
\begin{split}
&H_{AB} = g \sum_{i, \alpha, \beta} \vec{S}_i \cdot (a_{i,c_{\alpha}, \alpha} ^ {\dagger} \vec{\sigma}_{\alpha, \beta} a_{i,c_{\beta},\beta}),
\end{split}
\end{equation}
where $\vec{S}_i $ is the spin of the magnetic ion and $g>0$ is the exchange coupling. Note that here with a slight abuse of notation, the sum over index $i$ runs over sites of the honeycomb lattice, and basically each ion is coupled to three neighboring superconducting sites. Here, the color index should be understood as $c_{\uparrow}=c$ and $c_{\downarrow}=\bar{c}$, where $\bar{c}$ denotes a cyclic color operator, i.e., $\bar{r} = g, \bar{g} = b, \bar{b} = r$. In terms of Majorana fermions, the coupling Hamiltonian \eqref{kondo_majorana} reads as,
 
\begin{equation}\label{kondo_majorana}
\begin{split}
H_{AB} = \frac{g}{4} \sum_{i, c}&\Big[ S_i^x\left(i\gamma_{i, c, \uparrow}^- \gamma_{i, \bar{c}, \downarrow}^+ - i\gamma_{i, c, \uparrow}^+ \gamma_{i, \bar{c}, \downarrow}^-\right) \\ & - S_i^y\left(i\gamma_{i, c, \uparrow}^- \gamma_{i, \bar{c}, \downarrow}^- + i\gamma_{i, c, \uparrow}^+ \gamma_{i, \bar{c}, \downarrow}^+\right) \\ & -S_i^z\left(i\gamma_{i, c, \uparrow}^+  \gamma_{i, c, \uparrow}^- - i\gamma_{i, \bar{c}, \downarrow}^+ \gamma_{i, \bar{c}, \downarrow}^-\right) \Big].
\end{split}
\end{equation}

The total system $A+B$ is described by $H=H_A+H_B+H_{AB}$. We are aiming to treat the coupling Hamiltonian perturbatively. Each part of the system is characterized by an energy scale: the parent system is gapped with energy $\Delta_A$, and the impurity ions are characterized by an energy scale $W_B$, which is assumed to be zero.  For perturbation theory to work, the energy scales have to satisfy $W_B \ll g \ll \Delta_A$ \cite{Hsieh2016}. Within this range of energy scales, the degrees of freedom of system $A$ can be integrated out, resulting in an effective ring exchange interaction between magnetic ions in system $B$. In the absence of coupling $g=0$, the total system is highly degenerate, and one has to employ degenerate perturbation theory using the projection $P = \sum_j |\psi_A \otimes \psi_{B, j} \rangle\langle \psi_A \otimes \psi_{B, j}|$ to low-energy states. We evaluate the effect of coupling \eqref{kondo_majorana} order by order until the first nontrivial exchange interactions between the magnetic spin operators $S^{\alpha}$ arise. The nontrivial term arises when in a given perturbation order, the multiplication of the terms in \eqref{kondo_majorana} along the bonds generate stabilizer operators appearing in the ground state \eqref{A_superconductor}, i.e., the terms that project $|\psi_A\rangle$ on itself. The first nontrivial spin interaction arises at the sixth order of perturbation in $H_{AB}$, see Appendix \ref{appEq7} for details, giving rise to the following effective ring exchange Hamiltonian,   
 
\begin{equation}\label{H_CC}
\begin{split}
H_{eff}^{(6)} & = \frac{63 g^6}{4^{10}} \sum_{\hexagon} \left(\prod_{v \in \hexagon}S_v^y + 2\prod_{v \in \hexagon}S^x_v \right),
\end{split}
\end{equation}
where the sum over $\hexagon$ runs over all hexagonal plaquettes, shown by light red, blue, and green colors in Fig.~\ref{superconductor_scheme}, and $v$ denotes the vertices of a hexagon made of magnetic impurities. While \eqref{H_CC} resembles that of the color code model with six-body exchange interaction, the connection can be more precise by a unitary transformation on the sites of the hexagonal lattice. The latter is a bipartite lattice, so let us make the identity transformation on one sublattice $\{S^x, S^y, S^z\}\rightarrow \{S^x, S^y, S^z\}$ and a $\pi$-rotation around the spin $z$-axis on another sublattice as $\{S^x, S^y, S^z\}\rightarrow \{-S^x,- S^y, S^z\}$. Doing so, the Hamiltonian \eqref{H_CC} becomes   

\begin{equation}\label{H_CC2}
\begin{split}
H_{eff}^{(6)} & =- \frac{63 g^6}{4^{10}} \sum_{\hexagon} \left(\prod_{v \in \hexagon}S_v^y + 2\prod_{v \in \hexagon}S^x_v \right).
\end{split}
\end{equation}
This Hamiltonian is nothing but the color code model with local $\mathbb{Z}_2 \times \mathbb{Z}_2$ gauge symmetry and its spectrum is topologically ordered\cite{Bombin2006} with nonvanishing topological entanglement entropy\cite{Kargarian2008}.

\section{Topological orders induced by time-reversal invariant Majorana Cooper boxes}\label{TRMCB_model}

The network of $p$-wave superconducting chains introduced in the preceding section, though induces $\mathbb{Z}_2 \times \mathbb{Z}_2$ topological order in the magnetic impurities and substantiates the principles for engineering topological orders which are richer than the $\mathbb{Z}_2$ toric code model studied in Ref.[\onlinecite{Hsieh2017}], seems to be experimentally challenging to be synthesized.     

In this section, we take a different path to induce both $\mathbb{Z}_2$ and $\mathbb{Z}_2 \times \mathbb{Z}_2$ topological orders in a network of Majorana bound states which is experimentally more feasible. The idea is to use the Majorana bound states in time-reversal invariant topological superconductors (TRITOPS)\cite{Qi2009, Zhang2013,Schnyder2008}; these superconductors belong to class D\RNum{3} with both particle-hole and time-reversal symmetries. The fermionic operators $\Gamma_{E, \sigma} \quad (\sigma = \uparrow, \downarrow)$ of a zero-energy state satisfies $\Gamma^{\dagger}_{0, \sigma} =\pm i\Gamma_{0, -\sigma}$. Hence, zero-energy modes contain two Majorana bound states, which form Kramers' doublets and are protected due to time-reversal symmetry. Such pairs can appear at the ends of a semiconductor wire proximitized to the surface of a superconductor as shown in Fig.~\ref{TRITOPS_QD_scheme}. We then couple the Majorana bound states of TRITOPS to a quantum dot with strong Hubbard interaction. In what follows, we first give a description of this hybridization, largely based on Ref.[\onlinecite{Camjayi2017}], and then design proper lattice models whose low-energy physics is topologically ordered.

\subsection{TRITOPS coupled to quantum dots: model Hamiltonian}\label{TRITOPS}
\begin{figure}[t]
\center
\includegraphics[width=\linewidth]{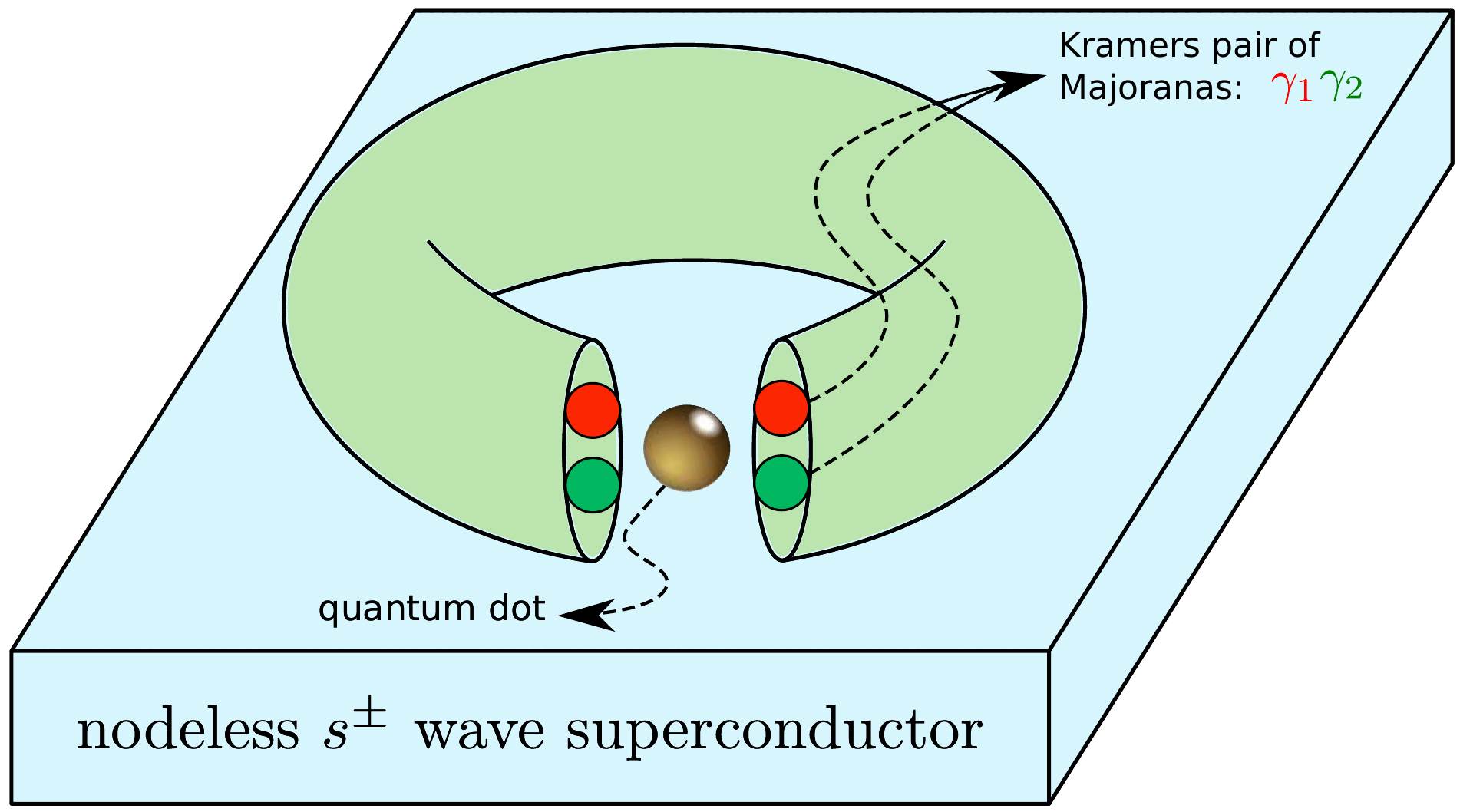}
\caption{A schematic of a time-reversal Majorana Cooper box (TRMCB). A semiconductor nanowire, light green wire, is in proximity to an $s^{\pm}$-wave superconductor. The system preserves the time-reversal symmetry, and pairs of Majorana bound states, shown as green and red solid dots, appear at the ends of nanowire forming Kramers' pairs. The bound states are tunnel coupled to a quantum dot shown by a solid brown sphere. \label{TRITOPS_QD_scheme}}
\end{figure}

There are several ways to engineer TRITOPS \cite{Nakosai2012, Deng2012, Wong2012, Zhang2013, Keselman2013}. One way is to use a hybrid system of 1D semiconductor with strong Rashba spin-orbit coupling proximitized to an $s^{\pm}$- wave iron-based superconductor \cite{Zhang2013} as shown in Fig.~\ref{TRITOPS_QD_scheme}. The system presents a time-reversal invariant version of Majorana Cooper boxes \cite{Beri2012}, and we refer to them as TRMCB. 

The quantum nanowire is described by the following Hamiltonian,

\begin{equation}
\begin{split}
H_{qw} = &\sum_{i=1}^N \sum_{\sigma}(-t c_{i+1, \sigma}^{\dagger} c_{i, \sigma} + i \lambda _{\sigma}c_{i+1, \sigma}^{\dagger} c_{i, \sigma} \\
&+ \Delta_{\sigma}c_{i+1, \sigma}^{\dagger} c_{i, \bar{\sigma}}^{\dagger} + \mathrm{H.c.} - \mu n_{i, \sigma}),
\end{split}
\end{equation}
where $\lambda_{\uparrow, \downarrow} = \pm \lambda$ is the Rashba spin-orbit coupling, $\Delta_{\uparrow, \downarrow} = \pm \Delta$ is the extended s-wave pairing, and $\mu$ is the chemical potential. The model preserves time-reversal and U(1) rotational symmetry. The latter symmetry can be broken by considering a more general form of Rashba coupling $i \lambda c_{i+1, \alpha}^{\dagger} \sigma^{x,y}_{\alpha, \beta} c_{i, \beta}$\cite{Keselman2013}. When $|\mu| < 2\lambda$, the superconductor is topological and pairs of Majorana bound states appear at the end of the chain as shown by green and red dots  in Fig.~\ref{TRITOPS_QD_scheme}. Pairs of bound states construct fermionic states described by operators $\Gamma_{L} = (\gamma_{L,\uparrow} - i \gamma_{L,\downarrow})/2$ for the left and $\Gamma_{R} = (\gamma_{R,\uparrow} + i \gamma_{R,\downarrow})/2$ for the right side. They can be either empty or occupied with no energy cost. In terms of localized wave function of Majorana bound states $\phi_{L/R}(x)$ and electron fields $\psi_{\sigma}^{\dagger}(x)$ of the continuum model, the fermionic zero modes can be written as\cite{Camjayi2017} 
\begin{equation}
\begin{split}
\Gamma_{L/R} = \int dx \phi_{L/R}(x)[\psi_{\uparrow}(x) \mp i\psi_{\downarrow}^{\dagger}(x)]. 
\end{split}
\end{equation} 

The Hamiltonian for the quantum dot is  
\begin{equation}\label{H_QD}
\begin{split}
H_d = \epsilon_d \sum_{\sigma} n_{d, \sigma} + U n_{d, \uparrow}n_{d, \downarrow}
\end{split}
\end{equation}
where $\epsilon_d$ is a gate-tunable energy level, $n_{d,\sigma}$ is spin-resolved level occupation and $U$ is onsite Hubbard interaction. 

Hybridization between the quantum dot and superconductor is described by the following tunneling Hamiltonian,  

\begin{equation}\label{coupling_QD_TRITOPS}
\begin{split}
H_{cd} = -t' \sum_{\sigma}(c_{L, \sigma}^{\dagger} + c_{R,\sigma}^{\dagger})d_{\sigma} + \mathrm{h.c.},
\end{split}
\end{equation}
 
By projecting tunneling Hamiltonian to the subgap space spanned by Majorana bound states and assuming that the quantum dot is at the particle-hole symmetric point, the effective coupling Hamiltonian reads as \cite{Camjayi2017} (see Appendix \ref{appEq13} for details):

\begin{equation}\label{Kondo_dot}
\begin{split}
H_{\mathrm{Kondo}} = \frac{4t'^2}{U}&\Big[ S_d^z \left( (n_L + n_R - 1) + i (\Gamma_L^{\dagger}\Gamma_R - \Gamma_R^{\dagger}\Gamma_L) \right) \\&+ i \left(S_d^-\Gamma_L^{\dagger}\Gamma_R^{\dagger} - S_d^+ \Gamma_R \Gamma_L\right)\Big],
\end{split}
\end{equation}   
 
In terms of Majorana operators, the above coupling Hamiltonian becomes

\begin{equation}\label{coupling-qdot-TRMCB}
\begin{split}
H_{\mathrm{Kondo}} = \frac{4t'^2}{U}&\Big[ S_d^x \left(-\frac{i}{2} ( \gamma_{R\uparrow}\gamma_{L\uparrow} + \gamma_{R\downarrow}\gamma_{L\downarrow} )\right) \\&+ S_d^y  \left(\frac{i}{2} (\gamma_{R\uparrow}\gamma_{L\downarrow} - \gamma_{R\downarrow}\gamma_{L\uparrow} )\right) \\&+ S_d^z  \left(\frac{i}{2} (\gamma_{L\uparrow}\gamma_{L\downarrow} - \gamma_{R\uparrow}\gamma_{R\downarrow})\right)\Big].
\end{split}
\end{equation}   

The terms containing the Majorana operators can be written in terms of spin operators in the presence of charging energy provided by a capacitor connected to the mesoscopic superconductor. In Ref.[\onlinecite{Fu2010}] it is argued that the charging energy term
\begin{equation}
\begin{split}\label{charging}
H_c(n) = \frac{1}{2C}(ne - Q_0)^2, \quad n = 0, \pm 1, \pm 2, \cdots, 
\end{split}
\end{equation}
 where charge $Q_{0}$ is set by the gate voltage, adds additional energy to the Hamiltonian. In the presence of charging energy \eqref{charging} the states with an  even and odd number of electrons will have different energies. The Majorana zero modes span a four-dimensional subspace $|\mathrm{even (odd)}, n_{L}, n_{R}\rangle$ with $n_{R},n_{L}=0,1$, i.e., $\{|2N, 0, 0\rangle, |2N + 1, 1, 0\rangle, |2N + 1, 0, 1\rangle, |2N + 2, 1, 1\rangle\}$. Tunning the gate voltage to $Q_0/e=2N+1$,  these states are no longer degenerate: odd-parity states $\{|2N+1, 1, 0\rangle, |2N+1, 0, 1\rangle\}$ have zero energy, while even-parity states $\{|2N, 0, 0\rangle, |2N+2, 1, 1\rangle\}$ cost charging energy $E_{\mathrm{ch}}=e^2/2C$. The even vs odd parities can be faithfully characterized by the electron number parity operator,  

\begin{equation}\label{parity_op}
\hat{P} = (-1)^{N_{\mathrm{Cooper~pairs}}} (1 - 2\Gamma_R^{\dagger}\Gamma_R)(1 - 2\Gamma_L^{\dagger}\Gamma_L),
\end{equation}
which can be written in terms of Majorana operators, $\hat{P} = (i\gamma_{L,\uparrow} \gamma_{L\downarrow}) (i\gamma_{R\uparrow} \gamma_{R\downarrow}) $. Therefore, the ground state subspace has a definite parity with $ \gamma_{L,\uparrow} \gamma_{L\downarrow}\gamma_{R\uparrow} \gamma_{R\downarrow}=+1$. As such, in this subspace the spin operators can be represented as  
\begin{equation}\label{pauli_majorana}
\begin{split}
&\sigma_x = -\frac{i}{2} \left( \gamma_{R\uparrow}\gamma_{L\uparrow} + \gamma_{R\downarrow}\gamma_{L\downarrow} \right), \\
&\sigma_y = \frac{i}{2} \left(\gamma_{R\uparrow}\gamma_{L\downarrow} - \gamma_{R\downarrow}\gamma_{L\uparrow} \right),\\
&\sigma_z =  \frac{i}{2} \left(\gamma_{L\uparrow}\gamma_{L\downarrow} - \gamma_{R\uparrow}\gamma_{R\downarrow}\right). 
\end{split}
\end{equation} 

Using this representation, the equation \eqref{coupling-qdot-TRMCB} is written as an effective spin exchange coupling 

\begin{equation}
\begin{split}\label{Kondo}
H_{\mathrm{Kondo}} = \frac{4t'^2}{U}\mathbf{S}_d\cdot \boldsymbol{\sigma}.
\end{split}
\end{equation}   

This expression shows that a TRMCB is Kondo coupled to a quantum dot in the strong interaction limit, where the local charge fluctuations are suppressed and the only remaining low-energy degree of freedom is spin-$1/2$ for our single orbital dot. In the next two subsections, we design a network of TRMCBs, each coupled to a quantum dot and show that how the topologically ordered states arise.     

\subsection{Inducing $\mathbb{Z}_2$ topological order}\label{Inducing_toric_code}
In the previous work in Ref.[\onlinecite{Hsieh2017}], the toric code model possessing $\mathbb{Z}_2$ topological order is induced in a system of superconducting chains (system A) Kondo coupled to magnetic ions (system B) using \eqref{coupling-qdot-TRMCB}. The parent system is formed of the Kitaev $p$-wave superconductor chains for spin-up electrons along the $x$-axis and for spin-down electrons along the $y$-axis, forming a square lattice version of the model shown in Fig.~\ref{superconductor_scheme}.

\begin{figure}[t]
\center
\includegraphics[width=\linewidth]{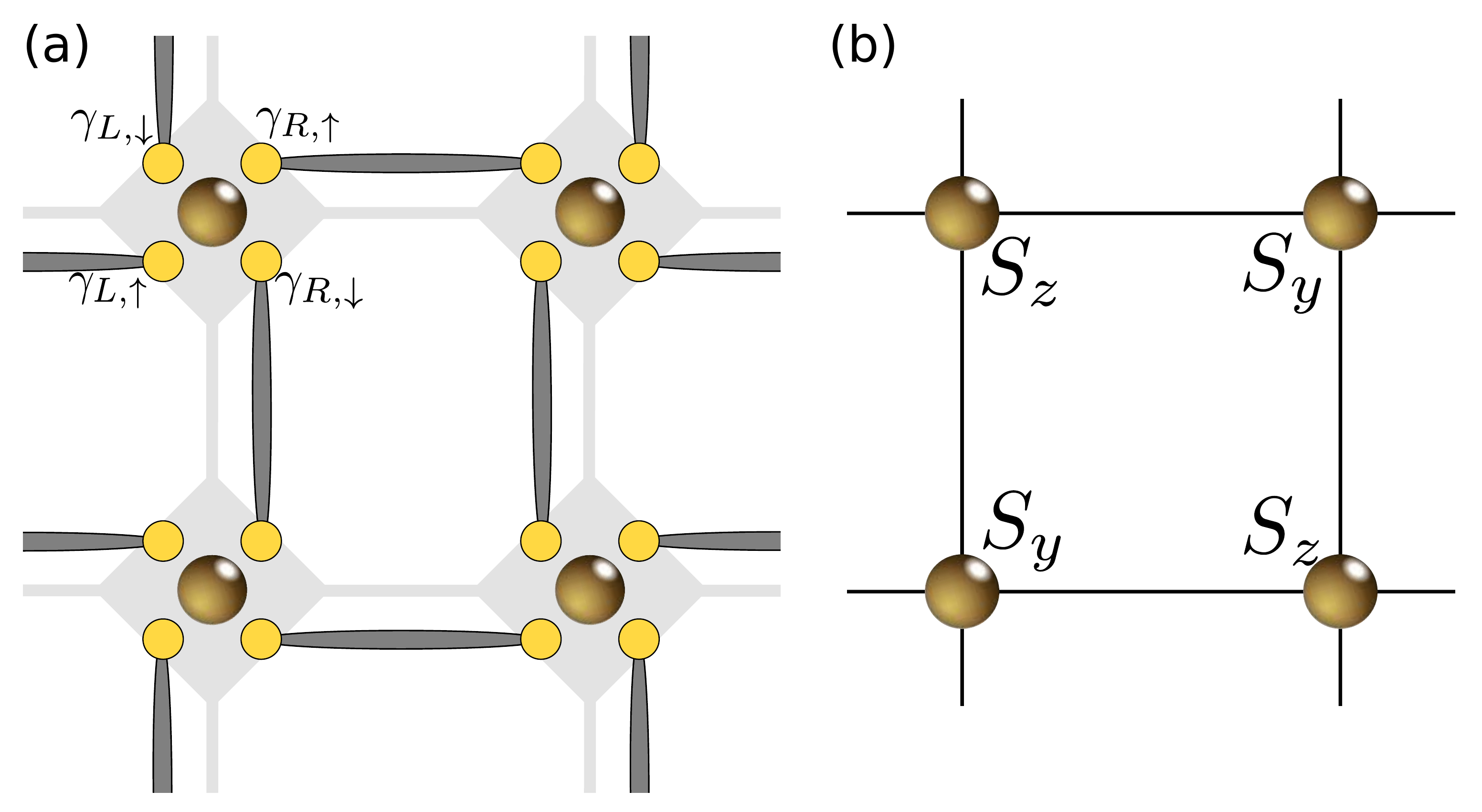}
\caption{(a) A network of TRMCBs on a square lattice. Majorana bound states are shown by solid yellow circles; there are four of them on each box, which are labeled as $\gamma_{L,\uparrow}, \gamma_{L,\downarrow}, \gamma_{R,\uparrow}, \gamma_{R,\downarrow}$. The boxes seen as grey diamonds belong to a superconducting substrate. The Majorana zero states belonging to neighboring boxes are connected to spin-polarized metallic leads. Each box is tunnel coupled to a quantum dot shown by a brown sphere. (b) In the limit of strong onsite Hubbard interaction, each dot is replaced by an effective magnetic moment. The square lattice of magnetic moments is described by a toric code model at the low-energy limit, where the plaquette operator is a product of spin operators shown. \label{toric_code_scheme}}
\end{figure}

Here, we aim to show that the same topological order can be inferred from a network of TRMCBs coupled to quantum dots via the hybridization mechanism explained in the  preceding subsection. In the limit of strong Hubbard interaction, the charge fluctuations on the quantum dot are suppressed giving rise to a spin-$1/2$ magnetic moment. The network of TRMCBs and quantum dots are shown in Fig.~\ref{toric_code_scheme}(a), where the Majorana fermions in TRMCBs are connected by spin-polarized metallic leads. The system is described by the following Hamiltonian,

\begin{equation}
\begin{split}
H = H_{c} + H_{\mathrm{lead}}+H_{\mathrm{tun}},
\end{split}
\end{equation}
where $H_{c}$ is the charging energy Hamiltonian in \eqref{charging}, which we write it as $H_{c}(N)=E_{\mathrm{ch}}(\hat{N}-N_g)^2$ with $\hat{N}$ as the total number operator of charges on the superconductor and $N_g$ is set by a gate voltage. $H_{\mathrm{lead}}$ describes electrons in leads and $H_{\mathrm{tun}}$ tunnel couples the leads to Majorana fermions to be described below. For concreteness, we first obtain an effective Hamiltonian describing a metallic lead coupled to Majorana modes at the ends. We assume that the lead with length $2D$ is coordinated as $-D\leq x\leq D$ and the electron operator with spin $\sigma$ is $\hat{\psi}_{\sigma}(x)=e^{ik_{F}x}\hat{\psi}_{R\sigma}(x)+e^{-ik_{F}x}\hat{\psi}_{L\sigma}(x)$, where R (L) denotes the right- (left-) moving electrons with Fermi wave vector $k_{F}$. A spin-polarized lead is described by 

\begin{equation}
\begin{split}
H^{\sigma}_{\mathrm{lead}} =-iv_{F}\int_{-D}^{D}dx\left(\hat{\psi}^{\dagger}_{R\sigma}(x)\partial_{x}\hat{\psi}_{R\sigma}(x)- \hat{\psi}^{\dagger}_{L\sigma}(x)\partial_{x}\hat{\psi}_{L\sigma}(x)  \right),
\end{split}
\end{equation} 
where $v_{F}$ is the Fermi velocity. The tunneling between the leads and Majorana modes $\gamma_{R,L}$ is given by

\begin{equation}
\begin{split}
H_{\mathrm{tun}}^{\sigma}=\sum_{j=R,L}t_{j}\hat{\psi}^{\dagger}_{j\sigma}e^{-i\hat{\chi}}\gamma_{j\sigma}+\mathrm{H.c.},
\end{split}
\end{equation}
where $\hat{\psi}^{\dagger}_{L\sigma}=\hat{\psi}^{\dagger}_{\sigma}(-D)$ and $\hat{\psi}^{\dagger}_{R\sigma}=\hat{\psi}^{\dagger}_{\sigma}(D)$ and $\hat{\chi}$ is the superconducting phase conjugated to the particle number $[\hat{\chi}, \hat{N}]=i$. $t_j$ is the tunneling matrix element between lead and Majorana bound states. We assume that the charging energy is the largest energy scale, i.e., $E_{c}> \Delta\gg t_{j}$ with $\Delta$ the induced superconducting gap. This allows us to integrate out the superconducting phase fluctuations. Using the Schrieffer-Wolff transformation for Majorana fermions coupled to leads\cite{Beri2012}, we obtain the following effective link Hamiltonian: 

\begin{equation}
\begin{split}\label{Heff}
H_{eff}^{\sigma}=H_{\mathrm{lead}}^{\sigma}-\sum_{l\neq j=L,R}\lambda^{+}_{lj}\hat{\psi}^{\dagger}_{l\sigma}\hat{\psi}_{j\sigma}\gamma_{l\sigma}\gamma_{j\sigma}-\sum_{j=L,R}\lambda^{-}_{jj}\hat{\psi}^{\dagger}_{j\sigma}\hat{\psi}_{j\sigma},
\end{split}
\end{equation}
 where $\lambda^{\pm}_{lj}=\left(\frac{1}{U^{+}}\pm\frac{1}{U^{-}} \right)t_{l}t_{j}$ with $U^{\pm}=H_c(N\pm1)-H_c(N)$. The second term describes the interaction between the Majorana states and electrons in the lead, and the third term merely shifts the chemical potential of the lead. A simple physical picture is obtained by treating this interaction using the mean-field theory: $\hat{\psi}^{\dagger}_{l\sigma}\hat{\psi}_{j\sigma}\gamma_{l\sigma}\gamma_{j\sigma}\simeq G^{\sigma}_{lj}\gamma_{l\sigma}\gamma_{j\sigma}+\Gamma^{\sigma}_{lj}\hat{\psi}^{\dagger}_{l\sigma}\hat{\psi}_{j\sigma}-G^{\sigma}_{lj}\Gamma^{\sigma}_{lj}$ with $G^{\sigma}_{lj}=\langle \hat{\psi}^{\dagger}_{l\sigma}\hat{\psi}_{j\sigma} \rangle $ and $\Gamma^{\sigma}_{lj}=\langle \gamma_{l\sigma}\gamma_{j\sigma}\rangle$. Using this decomposition, the Majorana part of the effective link Hamiltonian \eqref{Heff} reads as 

\begin{equation}
\begin{split}\label{H_Majorana1}
H^{\sigma}_{lj}=i \tilde{\lambda}^{\sigma}_{lj}\gamma_{l\sigma}\gamma_{j\sigma}, 
\end{split}
\end{equation}             
 where $\tilde{\lambda}^{\sigma}_{lj}=-2\mathrm{Im}(\lambda^{+}_{lj}G^{\sigma}_{lj})$. The total Majorana Hamiltonian is the sum over all links with $\sigma=\uparrow(\downarrow)$ for horizontal (vertical) leads. For simplicity and without loss of generality we assume that $\tilde{\lambda}^{\sigma}_{lj}=\tilde{\lambda}$ takes the same value for all leads. The total Majorana Hamiltonian then reads as  

\begin{equation}
\begin{split}\label{H_Majorana2}
H_{M}= \tilde{\lambda}\sum_{<l,j>\sigma} (i\gamma_{l\sigma}\gamma_{j\sigma}).  
\end{split}
\end{equation} 

This is the central result of this subsection. The above Hamiltonian is similar to the Kitaev superconducting chains \cite{Hsieh2017} and describes system A. Different terms commute with each other, and thus, the ground state can be written as 

\begin{equation}
\begin{split}\label{H_Majorana2}
|\psi_{A}\rangle= \prod_{<l,j>\sigma} (1-i\gamma_{l\sigma}\gamma_{j\sigma})|0\rangle, 
\end{split}
\end{equation} 
with energy $E_{A}=-N\tilde{\lambda}$, where $N$ is the total number of leads.
As prescribed in the previous subsection, each TRMBC is Kondo coupled to the magnetic ions, brown dots in Fig.~\ref{toric_code_scheme} forming system B, via the coupling $H_{\mathrm{Kondo}}$ in \eqref{Kondo}. Therefore, the low-energy spectrum of the total system is described by $H=H_{M}+H_{\mathrm{Kondo}}$. As shown in Ref.[\onlinecite{Hsieh2017}], a weak coupling treatment of the Kondo coupling within the perturbation theory induces a $\mathbb{Z}_2$ topological order described by the following magnetic Hamiltonian,

\begin{equation}
\begin{split}
H_{eff, B}=-K\sum_{i} S_{d}^{z}(\mathbf{r}_i)S_{d}^{y}(\mathbf{r}_i+\hat{x})S_{d}^{z}(\mathbf{r}_i+\hat{x}+\hat{y})S_{d}^{y}(\mathbf{r}_i+\hat{y}),
\end{split}
\end{equation}
where $K=5t'^8/U^4\tilde{\lambda}^3$ and $\mathbf{r}_i$ denotes the magnetic ions on the square lattice as shown in  Fig.~\ref{toric_code_scheme}(b). This is the  Wen's plaquette model with topological order \cite{Wen:PRL2003}. A simple unitary transformation on one sublattice, i.e., $S^{y}_d\rightarrow S^{z}_d$ and $S^{z}_d\rightarrow -S^{y}_d$, leads to the famous toric code Hamiltonian\cite{KITAEV20032}.    
 
\subsection{Inducing $\mathbb{Z}_2 \times \mathbb{Z}_2$ topological order}\label{inducing_colorcode_TRMCB}

\begin{figure}[t]
\center
\includegraphics[width=\linewidth]{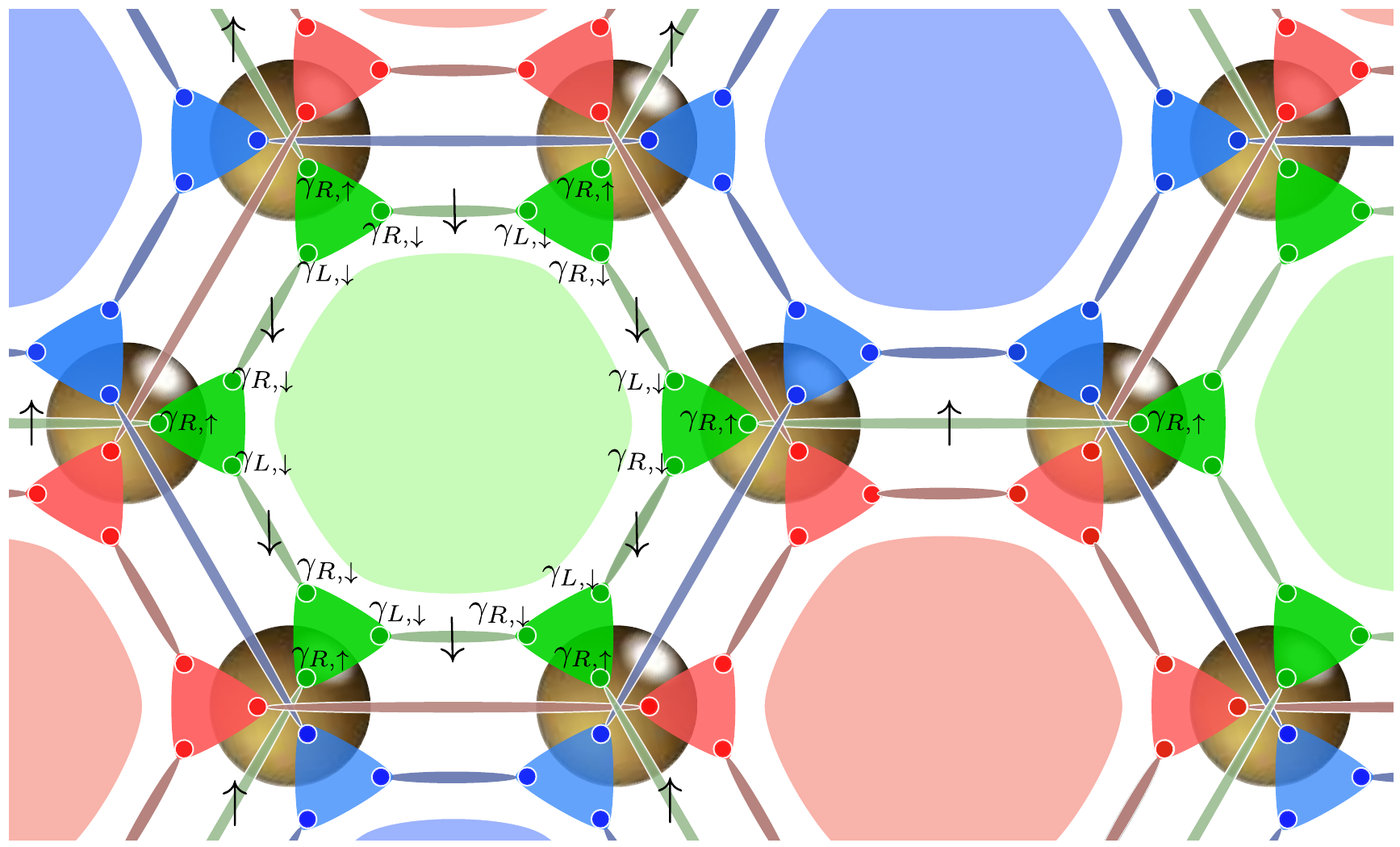}
\caption{A network of TRMCBs to induce $\mathbb{Z}_2 \times \mathbb{Z}_2$ topological order in magnetic impurities. Quantum dots reside on the vertices of the honeycomb lattice. Near each dot, there exist three TRMCBs shown by green, blue, and red colors. We use the same colors to show the Majorana bound states living on the corners of each box. Boxes with the same color, say green, are connected to each other by spin-polarized metallic leads, so we use the same color to highlight the corresponding leads. System A is then composed of three independent Majorana networks distinguished by their colors. On each site, three boxes are allowed to tunnel  to quantum dot, system B, which is described by the Kondo coupling in the limit of strong Hubbard interaction.\label{TRMCB_scheme}}
\end{figure}

The honeycomb lattice is the simplest lattice for the realization of $\mathbb{Z}_2 \times \mathbb{Z}_2$ topological order. We consider an arrangement of quantum dots on the vertices of a honeycomb lattice as shown in Fig.~\ref{TRMCB_scheme} with the assumption that there is no direct interaction between the effective spin of dots. The localized spin on the site is allowed to interact with three nearby TRMCBs distinguished by red, green, and blue colors. The boxes belong to a common superconducting substrate and are connected to each other by metallic leads. As discussed in preceding subsections, each TRMCB carries four Majorana bound states, of which we only consider $\gamma_{L,\downarrow}$, $\gamma_{R,\downarrow}$, and $\gamma_{R,\uparrow}$, and the fourth one remains dangling and it's not connected by metallic leads to neighboring boxes. Note that our final model is independent of the choice of Majorana states and, as shown in Fig.~\ref{TRMCB_scheme}, we show each box by a colored triangle with Majorana bound states sitting on the corners. The boxes with the same color are connected to each other by spin-polarized metallic leads. Six leads surrounding a hexagonal plaquette, say green, connect the Majorana bound states with spin down, and the out-going metallic leads with up spin connect Majorana $\gamma_{R,\uparrow}$ to the same Majorana on the neighboring green plaquettes. The same arrangement holds for other plaquettes with red and blue colors. In this way, we obtain three independent networks of Majorana states tunneling to metallic leads; they are clearly visible in  Fig.~\ref{TRMCB_scheme}.               

Following the same procedure lead to Eq.\eqref{H_Majorana1}, the low-energy description of Majorana states coupled to metallic leads for a given color, $c\in \{g,r,b\}$, is given by 
 
\begin{equation}
\begin{split}
H^{A}_c=\sum_{<lj>\in P_c}i\tilde{\lambda}\gamma^{c}_{l,\downarrow}\gamma^{c}_{j,\downarrow}+\sum_{<l\in P_{c}, j\in P_{c}'>} i\tilde{\lambda} \gamma^{c}_{l,\uparrow}\gamma^{c}_{j,\uparrow},
\end{split}
\end{equation}
where the first term originates from the hybridization of Majorana states with metallic leads within each plaquette $P_c$, and the second term describes the inter plaquette hybridization. The total Majorana Hamiltonian reads as $H_{M}=\sum_{c}H^{A}_{c}$ with the ground state $|\psi_{A}\rangle=\otimes_{c}|\psi^{c}_{A}\rangle$, where
 
\begin{equation}
\begin{split}
|\psi^{c}_{A}\rangle= \prod_{<l,j>\in P_c} (1-i\gamma^{c}_{l,\downarrow}\gamma^{c}_{j,\downarrow})\prod_{<l\in P_c,j\in P_c'>}(1-i\gamma^{c}_{l,\uparrow}\gamma^{c}_{j,\uparrow})|0\rangle, 
\end{split}
\end{equation} 
with energy $E_{A}=-3N\tilde{\lambda}$, where $N$ is the total number of leads. 

Each quantum dot is locally Kondo coupled to three adjacent TRMCBs as 
\begin{equation}
\begin{split}\label{Kondo_cc}
H_{\mathrm{Kondo}}=\frac{4t'^2}{U}\sum_{j,c}\mathbf{S}^{d}_j\cdot \boldsymbol{\sigma}^{c}_{j}, 
\end{split}
\end{equation} 
where the sum runs over lattice sites (quantum dots) of honeycomb lattice and three colors, and the Majorana representations of $\boldsymbol{\sigma}$ are defined in Eq.\eqref{pauli_majorana}. The topological order is induced on the lattice formed of magnetic moments of quantum dots, i.e., the moments $\mathbf{S}^{d}$ form system B, by treating the Kondo coupling \eqref{Kondo_cc} perturbatively in the total Hamiltonian $H=H_M+H_{\mathrm{Kondo}}$. Using the degenerate perturbation theory, the first nontrivial spin interaction terms appear at sixth order. The effective spin Hamiltonian is as follows (see Appendix \ref{appEq30} for details),       

\begin{equation}\label{HXZY}
\begin{split}
H_{eff,B}=-K\sum_{P}\left(X_{P} + Z_{P} - Y_{P}\right), 
\end{split}
\end{equation}
where the sum runs over all plaquettes of all colors, $K=63t'^{12}/4U^6\tilde{\lambda}^5$, and the plaquette operators read as  
\begin{equation}
\begin{split}
X_{P}=\prod_{j \in \hexagon}S_j^x, ~~Y_{P}=\prod_{j \in \hexagon}S_j^y, ~~Z_{P}=\prod_{j \in \hexagon}S_j^z.
\end{split}
\end{equation}
 
These plaquette operators form a set of stabilizers: they all commute with each other and square identity. Moreover, since $X_{P}Y_{P}=-Z_{P}$, the underlying induced topological order is governed by $\mathbb{Z}_2 \times \mathbb{Z}_2$ local gauge symmetry\cite{Bombin2006, Bombin2009, Kargarian_2010}.

\section{$\mathbb{Z}_2 \times \mathbb{Z}_2$ Majorana fermion codes}\label{constructing_Z2timesZ2_model}

Besides the spin models, topological order can also be realized in networks of Majorana fermions, known as Majorana fermion codes\cite{Bravyi2010}. A famous example is the realization of analogue of toric code model with $\mathbb{Z}_2$ topological order in an architecture pattern of superconducting islands hosting Majorana fermions \cite{Terhal:PRL2012, Plugge2016, Plugge2016_2, Plugge2016_3}. The basic building blocks of this architecture are Majorana Cooper boxes (MCBs)\cite{Beri2012} tunnel coupled to each other by metallic leads. Each MCB contains two nanowires with strong spin-orbit interaction (e.g., InAs or InSb) proximitized to a mesoscopic superconducting island \cite{Altland2013, Beri2013, Zazunov2014, Altland2014}. The wire becomes superconductor due to induced superconducting gap, and in a proper regime of Zeeman field and gate voltages the superconducting gap is topological with Majorana bound states appearing at the ends of wire \cite{Lutchyn2010, Alicea2012}. Each pair of Majorana bound states forms a single fermionic level which could be empty or occupied. Therefore, each island harbors four Majorana zero states spanning a four-dimensional local Hilbert space. In parallel to the arguments presented in Sec.\ref{TRITOPS}, the charging energy \eqref{charging} lifts the four-fold degeneracy to two-fold. Thus, in the restricted subspace of odd or even parity characterized by \eqref{parity_op} the island is effectively a spin-$1/2$ qubit \cite{Fu2010, Zazunov2011, Plugge2016, Plugge2016_2, Plugge2016_3,thomson2018, Tsvelik2020} which is described by the Pauli spin operators $\sigma^{\alpha} = i \epsilon_{\alpha\beta\delta}\gamma_\beta \gamma_\delta$ similar to \eqref{pauli_majorana}.       

\begin{figure*}[!htb]
\center
\includegraphics[width=\linewidth]{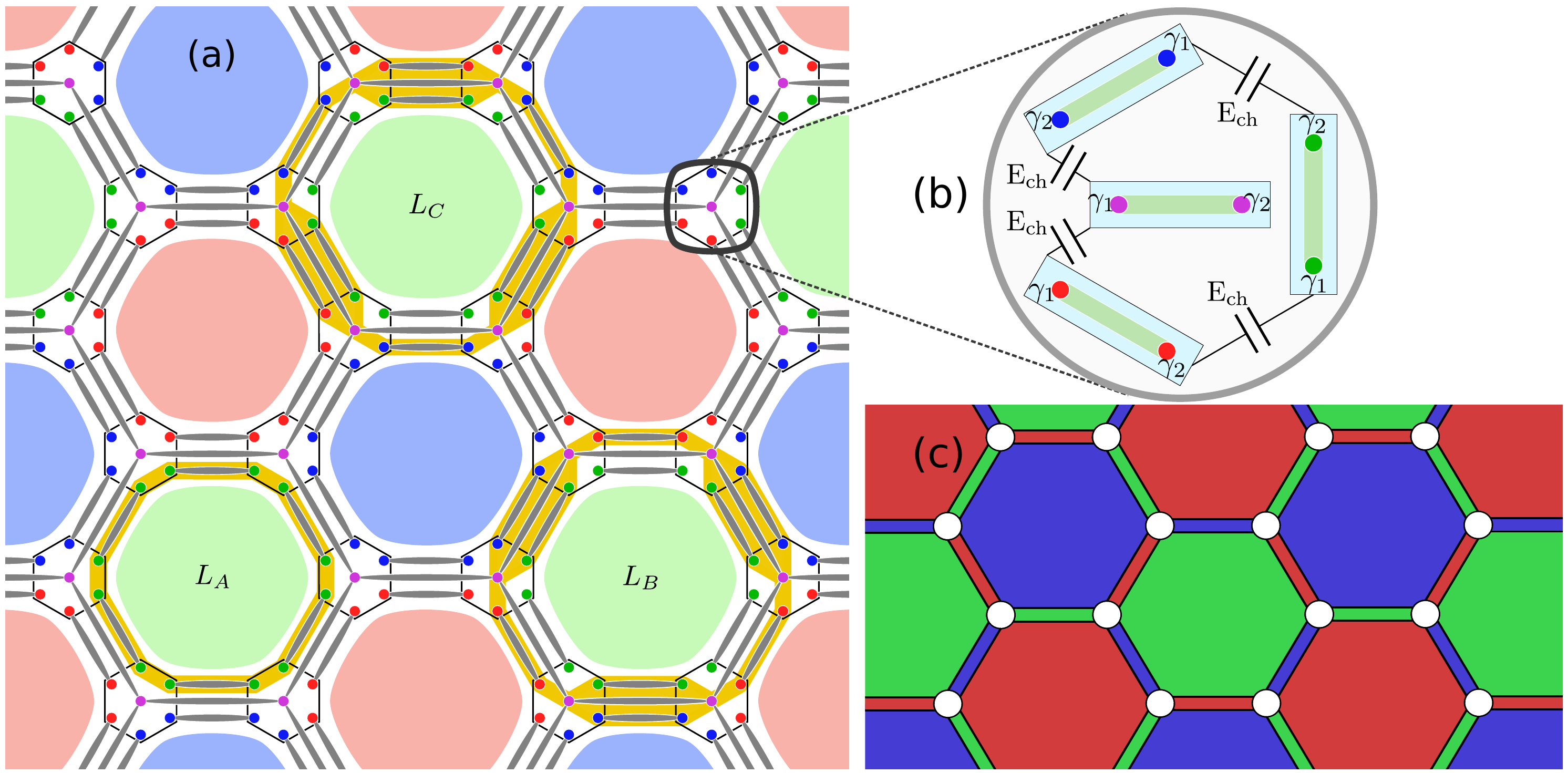}
\caption{An architecture pattern of colored Majorana Cooper boxes (CMCB) to construct a Majorana fermion code with $\mathbb{Z}_2 \times \mathbb{Z}_2$ topological order. (a) Each CMCB is shown by a solid hexagon and Majorana zero states sitting on the corners are labeled by red, green, and blue colors, and the one sitting on the center is colored purple. The plaquettes surrounded by Majoranas of the same color, say green, are shown by green and so on. For each plaquette $P$, three distinct string operators can be realized which are products of Majoranas encompassed. The strings are shown in yellow color and labeled by $L_A$, $L_B$, and $L_C$. In (b) a detailed structure of a CMCB island is displayed. Each island consists of four nanowires, light green thick lines, each in proximity to a superconductor shown by a rectangle. The charging energy is provided by connecting superconductors to capacitors with $E_{\mathrm{ch}}=e^2/C$. Note that each superconductor is also connected to a gate voltage by a capacitor not shown here. For each nanowire the pair of Majorana at the ends is shown by a color. (c) A piece of color code model, a model with $\mathbb{Z}_2 \times \mathbb{Z}_2$ topological order, emerged as a low-energy description of the Majorana-lead networks. The lattice is obtained by shrinking the CMCB islands in (a) into vertices. \label{MCB_scheme}}
\end{figure*}

To construct a Majorana fermion code with $\mathbb{Z}_2 \times \mathbb{Z}_2$ topological order we design an architecture pattern of Majorana bound states shown in Fig.~\ref{MCB_scheme}(a). Each solid hexagon represents a new colored Majorana Cooper box (CMCB) such that Majorana bound states are sitting on the corners where we use $r, b, g, p$ ($r\equiv \mathrm{red}, b\equiv \mathrm{blue}, g\equiv \mathrm{green}, p\equiv \mathrm{purple}$) colors to label them. Majoranas with the same colors belong to the ends of a nanowire proximitized to a superconductor. Each box consists of four topological nanowires. In Fig.~\ref{MCB_scheme}(b) a zoomed-out display of the box is shown, where the thick green lines are the nanowires and the underneath rectangle is a mesoscopic superconductor. The superconductors are connected by capacitors to each other to lift the degeneracy as discussed below. Also, each mesoscopic superconductor is connected to a gate voltage by a capacitor (not shown here). The boxes are arranged on the vertices of a hexagonal lattice and the Majorana bound states of neighboring boxes are tunnel coupled to each other by metallic leads shown by gray colors. Naturally, the plaquettes of the main lattice can be colored according to the color of the surrounding Majoranas as shown by light green, blue, and red in Fig.~\ref{MCB_scheme}(a). 

For an isolated box, see the window zoomed out in Fig.~\ref{MCB_scheme}(b), the uncoupled Majorana modes span a sixteen-fold degenerate subspace. The states can be represented as $|n_r, n_g, n_b, n_p\rangle$ where $n_c=0,1$ is the occupation of a pair of Majoranas. Coupling the superconductors to joint charging energy provided by the capacitors lifts the degeneracy. The Hamiltonian for a single isolated box is as 

\begin{equation}
\begin{split}\label{Hbox}
H_{\mathrm{box}}=E_{\mathrm{ch}}\sum_{j}\left(\hat{N}_j - N_{g}\right)^2 + E_{\mathrm{ch}}\sum_{<j,l>}\left(\hat{N}_{j}-\hat{N}_{l}\right)^2,
\end{split}
\end{equation}  
where the first term provides charging energy for each mesoscopic superconductor,  and by adjusting gate voltage to set $N_g=Q_{0}/e$ (see Eq. \eqref{charging}) to half-integers, a double degeneracy arises\cite{Fu2010}. The second term connects the superconductors to each other by capacitors, and the number operators read as $\hat{N}_c=2\hat{N}_{\mathrm{Cooper~pairs}}+\hat{n}_{c}$. One can immediately realize that the states which are fully empty or occupied, i.e., $\{|0000\rangle, |1111\rangle \}$, are the lowest degenerate states and the remaining states cost charging energy. The occupation state of each nanowire is characterized by the parity operator $\hat{P}_c=i\gamma_{c,1}\gamma_{c,2}$ where $\gamma_{c,1}$ and $\gamma_{c,2}$ form a single fermionic state. Let us define a joint parity operator between two fermionic states as a product of corresponding parities, $\hat{P}_{cc'}=\hat{P}_{c}\hat{P}_{c'}$. In the two-fold degenerate ground state space $P_{cc'}=+1$, hence, yielding an effective spin-1/2 representation of each box with the following Pauli operators:

\begin{equation}
\begin{split}\label{sigma_box}
&\sigma^x = \epsilon_{c, \bar{c}, \bar{\bar{c}}} \gamma_{\bar{\bar{c}},2}\gamma_{\bar{c},1}\gamma_{c,1} \gamma_{p,1}, \\
&\sigma^y = \epsilon_{c, \bar{c}, \bar{\bar{c}}}\gamma_{\bar{\bar{c}},2}\gamma_{\bar{c},1}\gamma_{c,2}\gamma_{p,1}, \\
&\sigma^z = i\gamma_{c,1}\gamma_{c,2}, \\
\end{split}
\end{equation} 
 where $\epsilon_{c, \bar{c}, \bar{\bar{c}}}$ (with $c=r,b,g$ with cyclic notation as before) is the Levi-Civita symbol. It is easy to see that the above representation satisfies the Pauli algebra $[\sigma^{\alpha},\sigma^{\beta}]=2i\epsilon_{\alpha\beta\gamma}\sigma^{\gamma}$. In the following, we will assume that $E_{\mathrm{ch}}$ in \eqref{Hbox} is the largest energy scale so that the description in \eqref{sigma_box} holds for all CMCBs.

We then obtain an effective description of CMCBs connected to each other by metallic leads by integrating over leads. A typical Majorana-lead coupling link between $\gamma_{1}$ and $\gamma_{2}$ is described by the following Hamiltonian:

\begin{equation}
\begin{split}
H_{12}=H_{\mathrm{lead}}+\sum_{j=1,2}H_{\mathrm{box}}^{(j)}+\sum_{j=1,2}\left(t_j\hat{\psi}^{\dagger}_{j}e^{-i\hat{\chi}_{j}}\gamma_j+ \mathrm{h.c.}\right),
\end{split}
\end{equation}
where $H_{\mathrm{box}}^{(j)}$ is the charging Hamiltonian in \eqref{Hbox} of CMCBs on both sides of a metallic lead. Upon integrating out the electron fields $\hat{\psi}^{\dagger}$ of the lead, the tunneling of Majorana bound states is obtained as\cite{Plugge2016_2}

\begin{equation}
\begin{split}
H_{12, \mathrm{tun}}=\sum_{j=1,2}H_{\mathrm{box}}^{(j)}+t_{12}e^{i(\hat{\chi}_{2}-\hat{\chi}_{1})}\gamma_2\gamma_{1}+ \mathrm{h.c.},
\end{split}
\end{equation}        
where $t_{12}=-t_1t_2^{*}G(1,2)/2$ with $G(1,2)$ as the electron propagator. Taking into account the same process for all links, the network shown in Fig.~\ref{MCB_scheme}(a) is modeled as 

\begin{equation}
\begin{split}\label{H_CMCB}
H=\sum_{j}H_{\mathrm{box}}^{(j)}+\sum_{<j,l>} t_{lj}e^{i(\hat{\chi}_{j}-\hat{\chi}_{l})}\gamma_j\gamma_{l}+ \mathrm{h.c.}. 
\end{split}
\end{equation} 

In this model, the charging energy $E_{\mathrm{ch}}$ is the largest energy scale so that the superconducting islands are in the ground state subspace, and this allows us to treat the tunneling term perturbatively. Starting from the ground states of two neighboring CMBCs, the second term in \eqref{H_CMCB} describes a process in which a charge is transferred from one island to another one, hereby exciting both islands due to charging energy terms. Therefore, by projecting to the ground state subspace, the first-order perturbation vanishes as it excites two neighboring boxes. At the second order,  the reversing of the hopping process returns the islands to the ground state, but it is an overall shift in energy and is therefore discarded. At the same order,  other terms, e.g., by taking two parallel links connecting two islands, also appear,  which vanish upon projecting to the ground state subspace. For instance consider two islands, $j$ and $l$, and Majorana states $\gamma_{r,j}$, $\gamma_{g,j}$ belonging to island $j$ and $\gamma_{r,l}$, $\gamma_{g,l}$ to island $l$. Suppose the islands are in states $|0000\rangle_{j}\otimes |0000\rangle_{l}$ prior the coupling to each other by metallic leads. The tunneling term $e^{i(\hat{\chi}_{j}-\hat{\chi}_l)}\gamma_{r,j}\gamma_{r,l}$ excites the islands to $|1000\rangle_{j}\otimes |1000\rangle_{l}$, and the tunneling of parallel lead $e^{i(\hat{\chi}_{l}-\hat{\chi}_j)}\gamma_{g,l}\gamma_{g,j}$ excites them to $|1100\rangle_{j}\otimes |1100\rangle_{l}$, which vanishes by projecting to the ground state subspace.    

By inspection, one can realize that the first nontrivial terms appear at sixth and twelfth orders of perturbation. Here a combination of tunneling processes around a plaquette returns the islands to their ground states. For each plaquette, three minimal closed loops can be realized, and we label them as $L_{A}, L_{B}, L_{C}$ as shown in Fig.~\ref{MCB_scheme}(a) by yellow color. For example, for a green plaquette, the $L_A$ loop contains only green Majorana states surrounding the plaquette, and the $L_{B}$ and $L_C$ contain all colored Majorana of each island in an ordered shown. Corresponding to each loop, we define a closed string operator as a product of Majorana operators encompassed, i.e., for a plaquette $P$ they are given by

\begin{equation}
\begin{split}\label{Loops}
Z_{P}=-\prod_{j\in L_{A}}\gamma_{j,c}, ~~X_{P}=\prod_{j\in L_{B}}\gamma_{j}, ~~Y_{P}=\prod_{j\in L_{C}}\gamma_{j}. 
\end{split}
\end{equation} 

The plaquette operators in \eqref{Loops} square identity individually and commute with each other mutually: $Z_{P}^2=X_{P}^2=Y_{P}^2=1$ and $[Z_{P}, X_{P}]=[Z_P,Y_{P}]=[X_{P},Y_{P}]=0$, and since $X_{P}Y_{P}Z_{P}=-1$, they form a local $\mathbb{Z}_2 \times \mathbb{Z}_2$ gauge group. 
The low-energy effective description of \eqref{H_CMCB} is given by plaquette operators \eqref{Loops} arising at sixth and twelfth orders of perturbations (see the details in Appendix \ref{appEq38}),

\begin{equation}
\begin{split}\label{CC_H}
H_{eff}=K^{(6)} \sum_{P} Z_{P}+ K^{(12)} \sum_{P} \left(X_{P}+Y_{P}\right),
\end{split}
\end{equation} 
where the first sum runs over all closed strings $L_A$ and the second sum runs over  $L_B$ and $L_C$ in \eqref{Loops}. The coefficients read as  $K^{(6)}=\kappa t^6/E_{\mathrm{ch}}^5$ and $K^{(12)}=-\kappa' t^{12}/E_{\mathrm{ch}}^{11}$ with numerical factors $\kappa,\kappa'>0$. 

The correspondence of the model in \eqref{H_CMCB} with the color code model is shown in Fig.~\ref{MCB_scheme}(c). By shrinking each CMCB in Fig.~\ref{MCB_scheme}(a) to a vertex, a hexagonal lattice is obtained. The correspondence can be made more concrete by noticing that the operators in \eqref{Loops} can be written as a product of corresponding Pauli operators of CMCBs around a plaquette using the identifications in \eqref{sigma_box} in the low-energy sector. The operators in \eqref{Loops} are   
\begin{equation}
\begin{split}
X_{P}=\prod_{j \in \hexagon}\sigma_j^x, ~~Y_{P}=\prod_{j \in \hexagon}\sigma_j^y, ~~Z_{P}=\prod_{j \in \hexagon}\sigma_j^z.
\end{split}
\end{equation}

These operators form the stabilizer group of the color code model, a piece of the lattice is shown in Fig.~\ref{MCB_scheme}(c), described by the effective Hamiltonian in \eqref{CC_H}. Using a unitary transformation, similar to what leads to \eqref{H_CC}, the ground state(s) of the model \eqref{CC_H} falls into a vortex free sector where $X_{P}|\psi_0\rangle=Z_{P}|\psi_0\rangle=|\psi_0\rangle$ for all stabilizer operators.  Indeed, the effective Hamiltonian \eqref{CC_H} describes a $\mathbb{Z}_2 \times \mathbb{Z}_2$ topologically ordered system in a purely Majorana model. To show this, we note that for a hexagonal lattice embedded on a closed surface, e.g., a torus, the following relations hold:    
\begin{equation}
\begin{split}\label{stablizer}
\prod_{r,b}X_rX_b=1,~~\prod_{r,g}Z_rZ_g=1,~~\prod_{b,g}Y_bY_g=1. 
\end{split}
\end{equation}

This shows that not all plaquette operators are independent. Assuming a lattice with $N$ plaquettes, the constraints in \eqref{stablizer} imply that the ground state subspace in the Hilbert space is $2^{2N}/2^{2N-4}=2^4$ fold degenerate, a hallmark of the color code model \cite{Bombin2006}.

\section{Conclusions\label{conclusions}}
The chief goal of this paper is to introduce models of Majorana fermions whose low-energy sector is topologically ordered with underlying $\mathbb{Z}_2 \times \mathbb{Z}_2$ and $\mathbb{Z}_2$ local gauge symmetry. The basic idea is to design platforms made of conventional ingredients in ways that the interplay between energy scales leads to exotic low-energy excitations. One approach is to use one-dimensional topological superconductors, e.g., the Kitaev chains, with both spin species and Kondo couple them to a lattice of free magnetic moments, a bootstrapping way of inducing topological orders in magnetic systems\cite{Hsieh2017}. 

In all designs it is important to have well-localized Majorana bound states. In $s_{\pm}$ wave superconductors, Majorana zero modes are localized on a length scale of order of coherence length $\xi = 100$ nm \cite{Scheurer2015}, so we can ignore the direct tunneling between them using nanowires of order $1\mu m$ in length. Mesoscopic superconductor charging energy is $E_{\mathrm{ch}} = 1$ meV \cite{Plugge2016} and tunneling amplitude is $t \ll E_{\mathrm{ch}}$. For $t \gtrsim 0.33E_{\mathrm{ch}}$, the ground state of the system is a frustrated 2D Ising phase which is not useful for quantum information purposes \cite{Terhal:PRL2012}, leading us to consider tunneling amplitude $t \lesssim 0.3 E_{\mathrm{ch}}$. Also, the pairing gap should be large enough for Majoranas to be well separated from the quasiparticle continuum. The estimated gap size of the iron base superconductor at $5$ K is about $3.6$ meV \cite{Sato2008} with the induced gap of order of $\Delta_{ind} \approx 0.46$ meV, so the effective energy scale in Majoana system is $\tilde{\lambda} \lesssim 0.09 E_{\mathrm{ch}}$ .

We introduce a network of one-dimensional superconductors coupled to a hexagonal lattice of magnetic ions. Performing a weak-coupling perturbation theory indicates that the magnetic system acquires a $\mathbb{Z}_2 \times \mathbb{Z}_2$ topological order. Our findings and the construction of $\mathbb{Z}_2$ topological order in Ref.[\onlinecite{Hsieh2017}] provide a conceptual framework on how to design exotic phases in more accessible settings. To connect the latter models to even more experimentally accessible settings, we first showed that a heterostructure of semiconductor nanowires proximitized to conventional superconductors, which by now have been experimentally realized thanks to almost a decade of tremendous efforts to produce heterostructures with high quality and finding some evidence of Majorana fermions at the end of nanowires (see Ref.[\onlinecite{Lutchyn:NatRev2018}] and references therein), can be designed in a way to induce $\mathbb{Z}_2$ topological order in a square lattice of magnetic ions which is experimentally more accessible than the construction based on the Kitaev chains. We then extend the model to magnetic ions arranged on the vertices of a honeycomb lattice and show that a $\mathbb{Z}_2 \times \mathbb{Z}_2$ topological order is induced on the magnetic sublattice. 

Also, motivated by recent works showing that the topological quantum computations can also be performed on quantum hardware made of only Majorana fermions, we introduce a network of Majorana bound states connected to each other by metallic leads. We first introduce a new Majorana Cooper box made of four nanowires, in contrast to two nanowires used in prior works. Charging energies provided by capacitors bridged the superconductors yield an effective spin-1/2 representation of boxes. The islands are architectured on a honeycomb lattice and are connected to each other by metallic leads allowing charges to tunnel between boxes. We show that the model at low energy has a local $\mathbb{Z}_2 \times \mathbb{Z}_2$ gauge structure which is an exact symmetry of the system. At the weak tunneling limit, the effective low-energy Hamiltonian describes a $\mathbb{Z}_2 \times \mathbb{Z}_2$ topological order. Hence, our model is a Majorana surface code with enriched topological order and maybe potentially useful for quantum computations and gate implementations with enhanced capabilities which may merit attention.

\section{ Acknowledgments}
The authors would like to acknowledge the support from Sharif University of Technology.

\appendix
\begin{widetext}
\section{Derivation of Eq.\eqref{H_CC}\label{appEq7}}
Using the degenerate perturbation theory, the effective Hamiltonian reads as

\begin{align}\label{effective_H}
H_{eff} = H_0 + P H_{AB} \sum_{n = 0}^\infty u^n P, \quad u = \frac{1}{E^0 - H_0} P' H_{AB},
\end{align}
where $P$ is the projection to the groundstate subspace and $P'=1-P$. 
The first-order term $H_{eff}^{(1)}  = P H_{AB} P $ vanishes because $H_{AB}|\psi_A^{gs}\rangle = |\psi_A^{ex}\rangle$ resulting in $ \langle\psi_A^{gs}|H_{AB}|\psi_A^{gs}\rangle = 0$.

Second-order effective Hamiltonian only shifts energy by a constant value $N g^2/16$. The first nontitvial term should be a multiplication of stabilizer operators living on the links of the lattice. Odd orders are always zero. Fourth order effective Hamiltonian adds a constant value to energy. Sixth-order perturbation is the first nontrivial one and contains three terms. Here, we write $\gamma_{i, c, \sigma}$ as $\gamma_{c, \sigma}(\mathbf{r_i})$. The first term is:

\begin{equation}\label{sixth_order_answer}
\begin{split}
H_{eff, 1}^{(6)} & = -\left(\frac{g}{4}\right)^6 \frac{63}{4^4}
\Big(\sum_j |\psi_A^{gs} \otimes \psi_{B, j} \rangle\langle \psi_A^{gs} \otimes \psi_{B, j}|\Big)
\sum_{c, \mathbf{r}_i}
\Big(-iS^y(\mathbf{r}_i) \gamma_{\bar{c},\uparrow}^+(\mathbf{r}_i) \gamma_{\bar{\bar{c}}, \downarrow}^+(\mathbf{r}_i)\Big)
\Big(-iS^y(\mathbf{r_i +‌ \delta_1}) \gamma_{\bar{c}, \uparrow}^-(\mathbf{r_i +‌ \delta_1}) \gamma_{\bar{\bar{c}}, \downarrow}^-(\mathbf{r_i +‌ \delta_1})\Big)\\
& \times\Big(-iS^y(\mathbf{r_i +‌ \delta_1 - \delta_3}) \gamma_{\bar{c}, \uparrow}^+(\mathbf{r_i +‌ \delta_1 - \delta_3}) \gamma_{\bar{\bar{c}}, \downarrow}^+(\mathbf{r_i +‌ \delta_1 - \delta_3})\Big)\\
&\times\Big(-iS^y(\mathbf{r_i +‌ \delta_1 - \delta_3 + \delta_2}) \gamma_{\bar{c}, \uparrow}^-(\mathbf{r_i +‌ \delta_1 - \delta_3 + \delta_2}) \gamma_{ \bar{\bar{c}}, \downarrow}^-(\mathbf{r_i +‌ \delta_1 - \delta_3 + \delta_2})\Big)\\
&\times\Big(-iS^y(\mathbf{r_i - \delta_3 + \delta_2}) \gamma_{\bar{c}, \uparrow}^+(\mathbf{r_i - \delta_3 + \delta_2}) \gamma_{\bar{\bar{c}}, \downarrow}^+(\mathbf{r_i - \delta_3 + \delta_2})\Big)
\Big(-iS^y(\mathbf{r_i + \delta_2}) \gamma_{\bar{c}, \uparrow}^-(\mathbf{r_i + \delta_2}) \gamma_{\bar{\bar{c}}, \downarrow}^-(\mathbf{r_i + \delta_2})\Big) \\
& \times \Big(\sum_{j '} |\psi_A^{gs} \otimes \psi_{B, j'} \rangle\langle \psi_A^{gs} \otimes \psi_{B, j'}|\Big),
\end{split}
\end{equation}
where $\mathbf{r}_i$ is the upper-left site on a plaquette with color $c$ and $\mathbf{\delta_1} = (a, 0), \mathbf{\delta_2} = (-a/2, \sqrt{3}a/2), \mathbf{\delta_3} = (-a/2, -\sqrt{3}a/2)$ are three neighbourings with lattice constant $a$. We rearrange this term as follows:

 \begin{equation}\label{sixth_order_answer}
\begin{split}
H_{eff, 1}^{(6)} & = -\left(\frac{g}{4}\right)^6 \frac{63}{4^4}
\Big(\sum_j |\psi_A^{gs} \otimes \psi_{B, j} \rangle\langle \psi_A^{gs} \otimes \psi_{B, j}|\Big)
\sum_{c, \mathbf{r}_i}
\Big(iS^y(\mathbf{r}_i) \gamma_{\bar{\bar{c}}, \downarrow}^+(\mathbf{r}_i) \gamma_{\bar{\bar{c}}, \downarrow}^-(\mathbf{r_i + \delta_2})\Big)
\Big(-iS^y(\mathbf{r_i +‌ \delta_1}) \gamma_{\bar{c},\uparrow}^+(\mathbf{r}_i) \gamma_{\bar{c}, \uparrow}^-(\mathbf{r_i +‌ \delta_1}) \Big)\\
& \times\Big(-iS^y(\mathbf{r_i +‌ \delta_1 - \delta_3})\gamma_{\bar{\bar{c}}, \downarrow}^+(\mathbf{r_i +‌ \delta_1 - \delta_3}) \gamma_{\bar{\bar{c}}, \downarrow}^-(\mathbf{r_i +‌ \delta_1}) \Big)\\
&\times\Big(-iS^y(\mathbf{r_i +‌ \delta_1 - \delta_3 + \delta_2})  \gamma_{\bar{c}, \uparrow}^+(\mathbf{r_i +‌ \delta_1 - \delta_3})  \gamma_{\bar{c}, \uparrow}^-(\mathbf{r_i +‌ \delta_1 - \delta_3 + \delta_2}) \Big)\\
&\times\Big(-iS^y(\mathbf{r_i - \delta_3 + \delta_2}) \gamma_{\bar{\bar{c}}, \downarrow}^+(\mathbf{r_i - \delta_3 + \delta_2}) \gamma_{ \bar{\bar{c}}, \downarrow}^-(\mathbf{r_i +‌ \delta_1 - \delta_3 + \delta_2})\Big)
\Big(-iS^y(\mathbf{r_i + \delta_2}) \gamma_{\bar{c}, \uparrow}^+(\mathbf{r_i - \delta_3 + \delta_2})  \gamma_{\bar{c}, \uparrow}^-(\mathbf{r_i + \delta_2}) \Big) \\
& \times \Big(\sum_{j '} |\psi_A^{gs} \otimes \psi_{B, j'} \rangle\langle \psi_A^{gs} \otimes \psi_{B, j'}|\Big).
\end{split}
\end{equation}

This combination of Majorana fermions are product of six stabilizer operators, which maps the ground state to itself in subsystem A. Thus, the first term of effective Hamiltonian induced in subsystem B is

\begin{equation}
\begin{split}
H_{eff,1,B}^{(6)} & = \frac{63}{4^{10}} g^6 \sum_{\hexagon} \prod_{v \in \hexagon}S_v^y .
\end{split}
\end{equation}
where the sum $\hexagon$ runs over all hexagonal plaquettes, and $v$ denotes the vertices of a hexagon made of magnetic impurities.

 The second term is

\begin{equation}
\begin{split}
H_{eff, 2}^{(6)} & =  -\frac{63}{4^{10}} g^6 
\Big(\sum_j |\psi_A^{gs} \otimes \psi_{B, j} \rangle\langle \psi_A^{gs} \otimes \psi_{B, j}|\Big)
\sum_{c,\mathbf{r_i}}
\Big(-iS^x(\mathbf{r_i}) \gamma^+_{\bar{\bar{c}} ,\uparrow}(\mathbf{r_i}) \gamma^-_{c, \downarrow}(\mathbf{r_i})\Big)\\
& \times \Big(iS^x(\mathbf{r_i} + \delta_1 - \delta_3 + \delta_2) \gamma^-_{\bar{\bar{c}}, \uparrow}(\mathbf{r_i} + \delta_1 - \delta_3 + \delta_2) \gamma^+_{c, \downarrow}(\mathbf{r_i} + \delta_1 - \delta_3 + \delta_2)\Big)\\
&\times\Big(-iS^x(\mathbf{r_i} - \delta_3 + \delta_2) \gamma^+_{\bar{\bar{c}}, \uparrow}(\mathbf{r_i} - \delta_3 + \delta_2) \gamma^-_{c, \downarrow}(\mathbf{r_i} - \delta_3 + \delta_2)\Big)
\Big(+iS^x(\mathbf{r_i} + \delta_1) \gamma^-_{\bar{\bar{c}}, \uparrow}(\mathbf{r_i} + \delta_{1}) \gamma^+_{c, \downarrow}(\mathbf{r_i} + \delta_{1})\Big)\\
&\times \Big(-iS^x(\mathbf{r_i} + \delta_1 - \delta_3) \gamma^+_{\bar{\bar{c}}, \uparrow}(\mathbf{r_i} + \delta_1 - \delta_3) \gamma^-_{c, \downarrow}(\mathbf{r_i} + \delta_{1} - \delta_{3})\Big)
\Big(+iS^x(\mathbf{r_i} + \delta_2) \gamma_{\bar{\bar{c}}, \uparrow}^-(\mathbf{r_i} + \delta_2) \gamma^+_{c, \downarrow}(\mathbf{r_i} + \delta_2)\Big)  \\
& \times \Big(\sum_{j '} |\psi_A^{gs} \otimes \psi_{B, j'} \rangle\langle \psi_A^{gs} \otimes \psi_{B, j'}|\Big),
\end{split}
\end{equation}
and we rearrange this term as
 
\begin{equation}
\begin{split}
H_{eff, 2}^{(6)} & =  -\frac{63}{4^{10}} g^6 
\Big(\sum_j |\psi_A^{gs} \otimes \psi_{B, j} \rangle\langle \psi_A^{gs} \otimes \psi_{B, j}|\Big)
\sum_{c,\mathbf{r_i}}
\Big(+iS^x(\mathbf{r_i}) \gamma^+_{\bar{\bar{c}} ,\uparrow}(\mathbf{r_i}) \gamma_{\bar{\bar{c}}, \uparrow}^-(\mathbf{r_i} + \delta_2) \Big)
\Big(+iS^x(\mathbf{r_i} + \delta_1 - \delta_3 + \delta_2) \gamma^+_{c, \downarrow}(\mathbf{r_i} + \delta_1 - \delta_3 + \delta_2) \gamma^-_{c, \downarrow}(\mathbf{r_i})\Big)\\
&\times\Big(+iS^x(\mathbf{r_i} - \delta_3 + \delta_2) \gamma^+_{\bar{\bar{c}}, \uparrow}(\mathbf{r_i} - \delta_3 + \delta_2)  \gamma^-_{\bar{\bar{c }}, \uparrow}(\mathbf{r_i} + \delta_1 - \delta_3 + \delta_2) \Big)
\Big(+iS^x(\mathbf{r_i} + \delta_1)  \gamma^+_{c, \downarrow}(\mathbf{r_i} + \delta_{1})\gamma^-_{c, \downarrow}(\mathbf{r_i} - \delta_3 + \delta_2)\Big)\\
&\times \Big(+iS^x(\mathbf{r_i} + \delta_1 - \delta_3) \gamma^+_{\bar{\bar{c}}, \uparrow}(\mathbf{r_i} + \delta_1 - \delta_3) \gamma^-_{\bar{\bar{c}}, \uparrow}(\mathbf{r_i} + \delta_{1}) \Big)
\Big(-iS^x(\mathbf{r_i} + \delta_2)  \gamma^+_{c, \downarrow}(\mathbf{r_i} + \delta_2) \gamma^-_{c, \downarrow}(\mathbf{r_i} + \delta_{1} - \delta_{3})\Big)\\
& \times \Big(\sum_{j '} |\psi_A^{gs} \otimes \psi_{B, j'} \rangle\langle \psi_A^{gs} \otimes \psi_{B, j'}|\Big). 
\end{split}
\end{equation}

Projecting to the groundstate of subsystem A, we obtain:

\begin{equation}
\begin{split}
H_{eff,2,B}^{(6)} & = \frac{63}{4^{10}} g^6 \sum_{\hexagon} \prod_{v \in \hexagon}S_v^x .
\end{split}
\end{equation}

The third term is 


\begin{equation}
\begin{split}
H_{eff, 3}^{(6)} & =  -\frac{63}{4^{10}} g^6 
\Big(\sum_j |\psi_A^{gs} \otimes \psi_{B, j} \rangle\langle \psi_A^{gs} \otimes \psi_{B, j}|\Big)
\sum_{\mathbf{r_i},c}
\Big(iS^x(\mathbf{r_i}) \gamma^+_{c, \uparrow}(\mathbf{r_i} + \delta_1 - \delta_3 + \delta_{2}) \gamma^-_{c,\uparrow}(\mathbf{r_i})\Big)
\Big(iS^x(\mathbf{r_i} + \delta_1) \gamma^+_{\bar{c}, \downarrow}(\mathbf{r_i}) \gamma^-_{\bar{c}, \downarrow}(\mathbf{r_i} + \delta_{1})\Big)\\
 &\times\Big(+iS^x((\mathbf{r_i} - \delta_3 + \delta_2))  \gamma^+_{c, \uparrow}(\mathbf{r_i} + \delta_{1}))  \gamma^-_{c, \uparrow}(\mathbf{r_i} - \delta_{3} + \delta_{2}) \Big)
\Big(iS^x(\mathbf{r_i} + \delta_2))  \gamma_{\bar{c}, \downarrow}^+(\mathbf{r_i} - \delta_{3} + \delta_{2})\gamma^-_{\bar{c}, \downarrow}(\mathbf{r_i} + \delta_{2})\Big)\\
&\times\Big(iS^x(\mathbf{r_i} + \delta_1 - \delta_3) \gamma^+_{c, \uparrow}(\mathbf{r_i} + \delta_{2}) \gamma^-_{c, \uparrow}(\mathbf{r_i} + \delta_{1} - \delta_{3}) \Big)\\
& \times \Big(-iS^x(\mathbf{r_i} + \delta_1 - \delta_3 + \delta_2)  \gamma^+_{\bar{c}, \downarrow}(\mathbf{r_i} + \delta_{1} - \delta_{3}) \gamma^-_{\bar{c}, \downarrow}(\mathbf{r_i} + \delta_{1} - \delta_{3} + \delta_{2})\Big)\\
& \times \Big(\sum_{j '} |\psi_A^{gs} \otimes \psi_{B, j'} \rangle\langle \psi_A^{gs} \otimes \psi_{B, j'}|\Big),
\end{split}
\end{equation}
which after projection, we obtain the following effective term acting on subsystem B: 

\begin{equation}
\begin{split}
H_{eff,3,B}^{(6)} & = \frac{63}{4^{10}} g^6 \sum_{\hexagon} \prod_{v \in \hexagon}S_v^x. 
\end{split}
\end{equation}

Therefore, up to sixth order obtain the following effective Hamiltonian in eq.\eqref{H_CC}: 

\begin{equation}
\begin{split}
H_{eff}^{(6)} & = \frac{63}{4^{10}} g^6 \sum_{\hexagon} \left(\prod_{v \in \hexagon}S_v^y + 2\prod_{v \in \hexagon}S^x_v \right),
\end{split}
\end{equation}

\section{Derivation of Eq.\eqref{Kondo_dot}}
\label{appEq13}
By projecting tunneling Hamiltonian $H_{cd}$ in eq.~\ref{coupling_QD_TRITOPS} to the low-energy regime we obtain

\begin{equation}
\begin{split}
H_c = t_{eff}\sum_{\sigma}(\Gamma_{L \sigma}^{\dagger} d_{\sigma} + d_{\sigma}^{\dagger} \Gamma_{R \sigma}) + h.c..
\end{split}
\end{equation} 

At the paricle-hole symmetric point, i.e., $U =  -\epsilon_d /2$ in eq.~\ref{H_QD}, the TRITOPS ground state energy is $E_0 = U/2$.  In a weak tunneling limit, $t_{eff} << E_0$, the tunneling Hamiltonian is considered as perturbation. First-order effective Hamiltonian is

\begin{equation}
\begin{split}
H_{eff}^{(1)} =  Q H_c Q = 0, 
\end{split}
\end{equation} 
where $Q = |\uparrow\rangle\langle \uparrow| + |\downarrow\rangle\langle \downarrow |$ is a projector to degenerate ground states subspace, and we define $P = |0\rangle\langle 0| + |\uparrow\downarrow\rangle\langle \uparrow\downarrow |$ as a projector to excited state subspace. Second-order perturbation is

\begin{equation}
\begin{split}
H_{eff}^{(2)} = &Q H_c \frac{1}{E_0 - H_0}  P H_c Q = \Big(|\uparrow\rangle\langle \uparrow| + |\downarrow\rangle\langle \downarrow| \Big) \Big(t_{eff}\sum_{\sigma}(\Gamma_{L \sigma}^{\dagger} d_{\sigma} + d_{\sigma}^{\dagger} \Gamma_{R \sigma}) + h.c.\Big) \left(\frac{1}{E_0 - H_0}\right) \Big(|0\rangle\langle 0| + |\uparrow\downarrow\rangle\langle \uparrow\downarrow| \Big) \\ &\times\Big(t_{eff}\sum_{\sigma}(\Gamma_{L \sigma}^{\dagger} d_{\sigma} + d_{\sigma}^{\dagger} \Gamma_{R \sigma}) + h.c.\Big) \Big(|\uparrow\rangle\langle \uparrow| + |\downarrow\rangle\langle \downarrow| \Big)  
\end{split}
\end{equation} 

\begin{equation}\label{H_eff}
\begin{split}
H_{eff}^{(2)} = & \frac{2t_{eff}^2}{U} \Big[\Big( \Gamma_{L,\uparrow}^{\dagger} \Gamma_{R,\uparrow} + \Gamma_{L,\uparrow}^{\dagger} \Gamma_{L, \uparrow} + \Gamma_{R,\uparrow}^{\dagger} \Gamma_{R,\uparrow} + \Gamma_{R,\uparrow}^{\dagger} \Gamma_{L,\uparrow} + \Gamma_{R,\downarrow} \Gamma_{L,\downarrow}^{\dagger} + \Gamma_{R,\downarrow} \Gamma_{R,\downarrow}^{\dagger} + \Gamma_{L,\downarrow} \Gamma_{L,\downarrow}^{\dagger} - \Gamma_{L,\downarrow} \Gamma_{R,\downarrow}^{\dagger} \Big) |\uparrow\rangle\langle \uparrow| \\
& - \Big(\Gamma_{R,\uparrow} \Gamma_{L,\downarrow}^{\dagger} + \Gamma_{R,\uparrow} \Gamma_{R, \downarrow}^{\dagger} + \Gamma_{L,\uparrow} \Gamma_{L,\downarrow}^{\dagger} + \Gamma_{L,\uparrow} \Gamma_{R,\downarrow}^{\dagger} - \Gamma_{L,\downarrow}^{\dagger} \Gamma_{L,\uparrow} - 	\Gamma_{L,\downarrow}^{\dagger} \Gamma_{L,\uparrow} - \Gamma_{R,\downarrow}^{\dagger} \Gamma_{R,\uparrow} - \Gamma_{R,\downarrow}^{\dagger} \Gamma_{L,\uparrow} \Big) |\downarrow \rangle\langle \downarrow| \\
& +\Big( -\Gamma_{L,\uparrow}^{\dagger} \Gamma_{R,\downarrow} - \Gamma_{L,\uparrow}^{\dagger} \Gamma_{L, \downarrow} - \Gamma_{R,\uparrow}^{\dagger} \Gamma_{R,\downarrow} - \Gamma_{R,\uparrow}^{\dagger} \Gamma_{L,\downarrow} + \Gamma_{R,\downarrow} \Gamma_{L,\uparrow}^{\dagger} + 	\Gamma_{R,\downarrow} \Gamma_{R,\uparrow}^{\dagger} - \Gamma_{L,\downarrow} \Gamma_{L,\uparrow}^{\dagger} + \Gamma_{L,\downarrow} \Gamma_{R,\uparrow}^{\dagger} \Big) |\downarrow\rangle\langle \uparrow| \\
& + \Big( \Gamma_{R,\uparrow} \Gamma_{L,\downarrow}^{\dagger} + \Gamma_{R,\uparrow} \Gamma_{R, \downarrow}^{\dagger} + \Gamma_{L,\uparrow} \Gamma_{L,\downarrow}^{\dagger} + \Gamma_{L,\uparrow} \Gamma_{R,\downarrow}^{\dagger} - \Gamma_{L,\downarrow}^{\dagger} \Gamma_{L,\uparrow} - 	\Gamma_{L,\downarrow}^{\dagger} \Gamma_{L,\uparrow} - \Gamma_{R,\downarrow}^{\dagger} \Gamma_{R,\uparrow} - \Gamma_{R,\downarrow}^{\dagger} \Gamma_{L,\uparrow} \Big) |\uparrow\rangle\langle \downarrow| \Big]
\end{split}
\end{equation} 

As mentioned in Sec.~\ref{TRMCB_model}, fermionic zero modes satisfy relation : $\Gamma_{L}^{\dagger} \equiv \Gamma_{L, \uparrow}^{\dagger} = i\Gamma_{L, \downarrow}$ and $\Gamma_{R}^{\dagger} \equiv \Gamma_{R, \uparrow}^{\dagger} = -i\Gamma_{R, \downarrow}$, so we can write Eq.~\ref{H_eff} in term of $\Gamma_{L}^{\dagger}$ and $\Gamma_{R}^{\dagger}$

\begin{equation}
\begin{split}
H_{\mathrm{Kondo}} = \frac{4t_{eff}^2}{U}\Big[ S_d^z \left( (n_L + n_R - 1) + i (\Gamma_L^{\dagger}\Gamma_R - \Gamma_R^{\dagger}\Gamma_L) \right) + i \left(S_d^-\Gamma_L^{\dagger}\Gamma_R^{\dagger} - S_d^+ \Gamma_R \Gamma_L\right)\Big],
\end{split}
\end{equation}  
where $n_{R(L)} = \Gamma_{R(L)}^{\dagger} \Gamma_{R(L)}$ is occupation number of $\Gamma_{R(L)}$ operator,  $S_z^d = (|\uparrow\rangle\langle\uparrow| - |\downarrow\rangle\langle\downarrow|) / 2$ is spin operator in $z$ direction, $S_-^d = |\downarrow\rangle\langle\uparrow|$ and $S_+^d = |\uparrow\rangle\langle\downarrow|$ are spin ladder operators.

\section{Derivation of Eq.\eqref{HXZY}}\label{appEq30}

Following the same procedure outlined in Appendix \ref{appEq7}, the sixth-order effective Hamiltonian is a ring exchange and can be written in term of stabilizer operators. In this order there are three ring exchanges for each plaquette of the honeycomb lattice. Each ring exchange belongs to one of Majorana networks which is distinguished by their colors. 
For a plaquette with color $c$, There are three nontrivial term at sixth order. First ring is composed of links with same color $c$

\begin{equation}
\begin{split}
H_{eff, 1}^{(6)} &= -\left(\frac{4t'^2}{U}\right)^6 \frac{63}{4^4} \frac{1}{\tilde{\lambda}^5}
\left(\sum_{j}|\psi_A^{gs}\otimes \psi_{B,j} \rangle \langle \psi_A^{gs}\otimes \psi_{B,j}|\right)
\sum_{i, c} S^x(\mathbf{r_i}) \left(\frac{i}{2}\gamma_{L, \downarrow}^{c}(\mathbf{r}_i) \gamma_{R,\downarrow}^c (\mathbf{r}_i + \hat{\delta}_2) \right) \\ 
& \times S^x(\mathbf{r}_i + \hat{\delta}_1) \left(\frac{i}{2} \gamma_{R, \downarrow}^c(\mathbf{r}_i)\gamma_{L, \downarrow}^c(\mathbf{r}_i + \hat{\delta}_1) \right)
S^x(\mathbf{r}_i + \hat{\delta}_1 - \hat{\delta}_3) \left(\frac{i}{2}\gamma_{L, \downarrow}^c(\mathbf{r}_i + \hat{\delta}_1 - \hat{\delta}_3) \gamma_{R, \downarrow}^c(\mathbf{r}_i + \hat{\delta}_1) \right)\\
& \times S^x(\mathbf{r}_i + \hat{\delta}_1 - \hat{\delta}_3 + \hat{\delta}_2) \left(\frac{i}{2} \gamma_{R, \downarrow}^c(\mathbf{r}_i + \hat{\delta}_1 - \hat{\delta}_3) \gamma_{L, \downarrow}^c(\mathbf{r}_i + \hat{\delta}_1 - \hat{\delta}_3 + \hat{\delta}_2)\right) \\
& \times S^x(\mathbf{r}_i - \hat{\delta}_3 + \hat{\delta}_2) \left(\frac{i}{2} \gamma_{L, \downarrow}^c(\mathbf{r}_i - \hat{\delta}_3 + \hat{\delta}_2) \gamma_{R, \downarrow}^c(\mathbf{r}_i + \hat{\delta}_1 - \hat{\delta}_3 + \hat{\delta}_2)  \right)
S^x(\mathbf{r}_i + \hat{\delta}_2) \left(\frac{i}{2} \gamma_{R, \downarrow}^c(\mathbf{r}_i - \hat{\delta}_3 + \hat{\delta}_2) \gamma_{L, \downarrow}^c(\mathbf{r}_i + \hat{\delta}_2)\right) \\
& \times \left( \sum_{j'}|\psi_A^{gs}\otimes \psi_{B,j'} \rangle \langle \psi_A^{gs}\otimes \psi_{B,j'}|\right).
\end{split}
\end{equation}
where $\mathbf{r}_i$ is the upper-left site of each plaquette with color $c$. After integrating out TRMCBs degrees of freedom, first term of  effective Hamiltonian  subspace will be

\begin{equation}
\begin{split}
&H_{eff, B,1}^{(6)} = -K  \sum_{P}X_p\\
&X_p \equiv \prod_{j \in \hexagon} S^x_j,
\end{split}
\end{equation}
where $K = 63t'^{12}/4U^6\tilde{\lambda}^5$. The next loop composed of $\bar{c}$ links is $H_{eff,B, 2}^{(6)} = -K \sum_{P} Z_p$, and the last loop is composed of  $\bar{\bar{c}}$ links and equals to $H_{eff,B, 3}^{(6)} = K  \sum_{P}Y_p$, where  $Z_p$ and $Y_p$ are

\begin{equation}
\begin{split}
Z_p \equiv \prod_{j \in \hexagon} S^z_j,
Y_p \equiv \prod_{j \in \hexagon} S^y_j.
\end{split}
\end{equation}

Thus, the effective Hamiltonian reads as eq.~\eqref{HXZY}:

\begin{equation}
\begin{split}
H_{eff, B} & = -K \sum_{P} \left(X_p + Z_p - Y_p\right)
\end{split}
\end{equation}

\section{Derivation of Eq.\eqref{CC_H}}\label{appEq38}

We define $P = P^{ch}_{gs}\otimes \sum_{j}|\psi_{MZM,j} \rangle \langle \psi_{MZM,j}|$ as projection operator to charging energy ground state, where $P^{ch}_{gs} = \prod_l \left( |0000\rangle_{ll} \langle 0000| + |1111\rangle_{ll} \langle 1111|\right)$ is product of the ground state subspace of $l$-th CMCB's charging energy, and $|\psi_{MZM,j}\rangle$ denotes a state in Majorana zero modes subspace. The first nonzero term appears at sixth order: 
 
\begin{equation}
\begin{split}
H_{eff}^{(6)}  = &-\frac{63}{4^4} \frac{t^6}{(4E_c)^5} P\Bigg(\sum_{i}  \Big(
e^{i(\hat{\chi}_c(\mathbf{r}_i)-\hat{\chi}_c(\mathbf{r}_i + \delta_1))}\gamma_{c,1}(\mathbf{r}_i) \gamma_{c,1}(\mathbf{r}_i + \delta_1)
e^{i(\hat{\chi}_c(\mathbf{r}_i + \delta_1)-\hat{\chi}_c(\mathbf{r}_i + \delta_1 - \delta_3))}\gamma_{c,2}(\mathbf{r}_i + \delta_1) \gamma_{c,2}(\mathbf{r}_i + \delta_1 - \delta_3) \\
 &\times e^{i(\hat{\chi}_c(\mathbf{r}_i + \delta_1 - \delta_3)-\hat{\chi}_c(\mathbf{r}_i + \delta_1 - \delta_3 + \delta_2))}\gamma_{c,1}(\mathbf{r}_i + \delta_1 - \delta_3) \gamma_{c,1}(\mathbf{r}_i + \delta_1 - \delta_3 + \delta_2)\\
& \times e^{i(\hat{\chi}_c(\mathbf{r}_i + \delta_1 - \delta_3 + \delta_2)-\hat{\chi}_c(\mathbf{r}_i + \delta_2 - \delta_3) )} \gamma_{c,2}(\mathbf{r}_i + \delta_1 - \delta_3 + \delta_2) \gamma_{c,2}(\mathbf{r}_i + \delta_2 - \delta_3) \\
& \times e^{i(\hat{\chi}_c(\mathbf{r}_i + \delta_2 - \delta_3)-\hat{\chi}_c(\mathbf{r}_i + \delta_2))}\gamma_{c,1}(\mathbf{r}_i + \delta_2 - \delta_3) \gamma_{c,1}(\mathbf{r}_i + \delta_2) 
e^{i(\hat{\chi}_c(\mathbf{r}_i + \delta_2)-\hat{\chi}_c(\mathbf{r}_i))}\gamma_{c,2}(\mathbf{r}_i + \delta_2) \gamma_{c,2}(\mathbf{r}_i) + h.c.\Big)\Bigg)P,
\end{split}
\end{equation} 
where $c$ is the color of each plaquette, $\mathbf{r}_i$ is upper-left site of $c$ plaquette. $e^{i\hat{\chi}_c(\mathbf{r}_i)}$ excite ith charging energy. For each $e^{i\hat{\chi}_c(\mathbf{r}_i)}$, there is one $e^{-i\hat{\chi}_c(\mathbf{r}_i)}$ which cancels this excitation. For example if $c = r$ we have 

\begin{equation}
\begin{split}
&P^{ch}_{gs}e^{i\hat{\chi}_r(\mathbf{r}_i)} e^{-i\hat{\chi}_r(\mathbf{r}_i)} P^{ch}_{gs} =  P^{ch}_{gs}e^{i\hat{\chi}_r(\mathbf{r}_i)} \left( |1000\rangle_{ii} \langle 0000| + |0111\rangle_{ii} \langle 1111| \right) \prod_{l \neq i} \left( |0000\rangle_{ll} \langle 0000| + |1111\rangle_{ll} \langle 1111| \right) \\
& = P^{ch}_{gs} \left( |0000\rangle_{ii} \langle 0000| + |1111\rangle_{ii} \langle 1111| \right) \prod_{l \neq i} \left( |0000\rangle_{ll} \langle 0000| + |1111\rangle_{ll} \langle 1111| \right) = P^{ch}_{gs} P^{ch}_{gs},
\end{split}
\end{equation}
so this combination of tunneling processes around a plaquette maps the islands to their ground states.  After integrating out charge degrees of freedom, effective Hamiltonian in Majorana Zero modes subspace will be

\begin{equation}
\begin{split}
H_{eff, MZM}^{(6)}  = &\frac{63}{4^4} \frac{2t^6}{(4E_c)^5} \sum_{i}
\left(i\gamma_{c,1}(\mathbf{r}_i) \gamma_{c,1}(\mathbf{r}_i + \delta_1)\right) 
\left(i\gamma_{c,2}(\mathbf{r}_i + \delta_1) \gamma_{c,2}(\mathbf{r}_i + \delta_1 - \delta_3)\right)
\left(i\gamma_{c,1}(\mathbf{r}_i + \delta_1 - \delta_3) \gamma_{c,1}(\mathbf{r}_i + \delta_1 - \delta_3 + \delta_2)\right)\\
& \times  \left(i\gamma_{c,2}(\mathbf{r}_i + \delta_1 - \delta_3 + \delta_2) \gamma_{c,2}(\mathbf{r}_i + \delta_2 - \delta_3)\right)
\left(i\gamma_{c,1}(\mathbf{r}_i + \delta_2 - \delta_3) \gamma_{c,1}(\mathbf{r}_i + \delta_2)\right) 
\left(i\gamma_{c,2}(\mathbf{r}_i + \delta_2) \gamma_{c,2}(\mathbf{r}_i)\right).
\end{split}
\end{equation} 

It can be written as follows
\begin{equation}
\begin{split}
H_{eff, A}^{(6)} = K^{(6)}\sum_{P} Z_P,~~~Z_P \equiv \prod_{j \in L_A}\gamma_j.
\end{split}
\end{equation}
where $K^{(6)} = 63t^6 / 2^9 E_c^5$, $L_A$ is the closed loop shown in Fig.~\ref{MCB_scheme}, and $P$ runs over each plaquette. 
If we rearrange $Z_P$ as

\begin{equation}
\begin{split}
Z_P  = &\Bigg(\sum_{i}
\left(i\gamma_{c,1}(\mathbf{r}_i) \gamma_{c,2}(\mathbf{r}_i)\right) 
\left(i\gamma_{c,1}(\mathbf{r}_i + \delta_1)\gamma_{c,2}(\mathbf{r}_i + \delta_1) \right)
\left(i\gamma_{c,1}(\mathbf{r}_i + \delta_1 - \delta_3) \gamma_{c,2}(\mathbf{r}_i + \delta_1 - \delta_3) \right)\\
& \times  \left(i \gamma_{c,1}(\mathbf{r}_i + \delta_1 - \delta_3 + \delta_2) \gamma_{c,2}(\mathbf{r}_i + \delta_1 - \delta_3 + \delta_2) \right)
\left(i\gamma_{c,1}(\mathbf{r}_i + \delta_2 - \delta_3) \gamma_{c,2}(\mathbf{r}_i + \delta_2 - \delta_3) \right) 
\left(i\gamma_{c,1}(\mathbf{r}_i + \delta_2) \gamma_{c,2}(\mathbf{r}_i + \delta_2)\right)
\Bigg),
\end{split}
\end{equation} 
it can be written in term of pauli operators (Eq.~\ref{sigma_box}) around the plaquette surrounding by $L_A$:

\begin{equation}
\begin{split}
&Z_P \equiv \prod_{j \in L_A}\sigma_j^z.
\end{split}
\end{equation}

The next nontrivial order is the one that maps each island from one ground state to the other one. For example

\begin{equation}
\begin{split}
P^{ch}_{gs}e^{\pm i\left(\hat{\chi}_r(\mathbf{r_i}) + \hat{\chi}_r(\mathbf{r_i}) + \hat{\chi}_r(\mathbf{r_i}) + \hat{\chi}_r(\mathbf{r_i})\right)}P^{ch}_{gs} 
= P^{ch}_{gs}  \left( |1111\rangle_{ii} \langle 0000| + |0000\rangle_{ii} \langle 1111| \right) \prod_{l \neq i} \left( |0000\rangle_{ll} \langle 0000| + |1111\rangle_{ll} \langle 1111| \right) = P^{ch}_{gs}P^{ch}_{gs}
\end{split}
\end{equation}

So if we excite the charge in one island, the island returns to its ground state via a  unneling process at twelfth order. The effective Hamiltonian has two nontrivial terms. The first one is

\begin{equation}
\begin{split}
H_{eff,B}^{(12)}  = &-\frac{\kappa'}{4} \frac{t^{12}}{E_c^{11}} P\Bigg(\sum_{i} \Big(
e^{i(\hat{\chi}_{\bar{\bar{c}}}(\mathbf{r}_i)-\hat{\chi}_{\bar{\bar{c}}}(\mathbf{r}_i + \delta_1))}\gamma_{\bar{\bar{c}},2}(\mathbf{r}_i) \gamma_{\bar{\bar{c}},2}(\mathbf{r}_i + \delta_1) \\
& \times e^{i(\hat{\chi}_c(\mathbf{r}_i)-\hat{\chi}_c(\mathbf{r}_i + \delta_1))} \gamma_{c,1}(\mathbf{r}_i) \gamma_{c,1}(\mathbf{r}_i + \delta_1)
e^{i(\hat{\chi}_p(\mathbf{r}_i)-\hat{\chi}_p(\mathbf{r}_i + \delta_1))} \gamma_{p,1}(\mathbf{r}_i) \gamma_{p,1}(\mathbf{r}_i + \delta_1)
e^{i(\hat{\chi}_{\bar{c}}(\mathbf{r}_i + \delta_1 - \delta_3) -\hat{\chi}_{\bar{c}}(\mathbf{r}_i + \delta_1))}  \gamma_{\bar{c},1}(\mathbf{r}_i + \delta_1 - \delta_3) \gamma_{\bar{c},1}(\mathbf{r}_i + \delta_1) \\
& \times e^{i(\hat{\chi}_{\bar{\bar{c}}}(\mathbf{r}_i + \delta_1 - \delta_3)-\hat{\chi}_{\bar{\bar{c}}}(\mathbf{r}_i + \delta_1 - \delta_3 + \delta_2))}\gamma_{\bar{\bar{c}},2}(\mathbf{r}_i + \delta_1 - \delta_3) \gamma_{\bar{\bar{c}},2}(\mathbf{r}_i + \delta_1 - \delta_3 + \delta_2) \\
& \times e^{i(\hat{\chi}_c(\mathbf{r}_i + \delta_1 - \delta_3)-\hat{\chi}_c(\mathbf{r}_i + \delta_1 - \delta_3 + \delta_2))} \gamma_{c,1}(\mathbf{r}_i + \delta_1 - \delta_3) \gamma_{c,1}(\mathbf{r}_i + \delta_1 - \delta_3 + \delta_2) \\ 
& \times e^{i(\hat{\chi}_p(\mathbf{r}_i + \delta_1 - \delta_3)-\hat{\chi}_p(\mathbf{r}_i + \delta_1 - \delta_3 + \delta_2))} \gamma_{p,1}(\mathbf{r}_i + \delta_1 - \delta_3) \gamma_{p,1}(\mathbf{r}_i + \delta_1 - \delta_3 + \delta_2)\\
& \times e^{i(\hat{\chi}_{\bar{c}}(\mathbf{r}_i + \delta_1 - \delta_3 + \delta_2) -\hat{\chi}_{\bar{c}}(\mathbf{r}_i + \delta_2 - \delta_3))} \gamma_{\bar{c}}(\mathbf{r}_i + \delta_2 - \delta_3) \gamma_{\bar{c},2}(\mathbf{r}_i + \delta_1 - \delta_3 + \delta_2) 
e^{i(\hat{\chi}_{\bar{\bar{c}}}(\mathbf{r}_i + \delta_2 - \delta_3)-\hat{\chi}_{\bar{\bar{c}}}(\mathbf{r}_i + \delta_2) )}\gamma_{\bar{\bar{c}},2}(\mathbf{r}_i + \delta_2 - \delta_3) \gamma_{\bar{\bar{c}},2}(\mathbf{r}_i + \delta_2)  \\
& \times e^{i(\hat{\chi}_c(\mathbf{r}_i + \delta_2 - \delta_3)-\hat{\chi}_c(\mathbf{r}_i + \delta_2) )}\gamma_{c,1}(\mathbf{r}_i + \delta_2 - \delta_3) \gamma_{c,1}(\mathbf{r}_i + \delta_2) 	
e^{i(\hat{\chi}_p(\mathbf{r}_i + \delta_1 - \delta_3)-\hat{\chi}_p(\mathbf{r}_i + \delta_1 - \delta_3 + \delta_2))} \gamma_{p,1}(\mathbf{r}_i + \delta_1 - \delta_3) \gamma_{p,1}(\mathbf{r}_i + \delta_1 - \delta_3 + \delta_2)\\
& \times e^{i(\hat{\chi}_{\bar{c}}(\mathbf{r}_i) -\hat{\chi}_{\bar{c}}(\mathbf{r}_i + \delta_2))} \gamma_{\bar{c},1}(\mathbf{r}_i) \gamma_{\bar{c},1}(\mathbf{r}_i + \delta_2)
+ h.c. \Big)\Bigg)P,
\end{split}
\end{equation} 
where $i$ is the upper-left site of a plaquette with color $c$ and $\kappa' > 0$ is a numerical factor. There is one $e^{\pm i\left(\hat{\chi}_r(\mathbf{r}_i) + \hat{\chi}_g(\mathbf{r}_i) + \hat{\chi}_b(\mathbf{r}_i) + \hat{\chi}_p(\mathbf{r}_i)\right)}$ operator for each CMCB around the plaquette. There is another term which includes  the purple tunneling. After integrating charges degrees of freedom both contributes the same: 

\begin{equation}
\begin{split}
H_{eff,B}^{(12)} = K^{(12)} \sum_{P} X_P,~~~ X_P = \prod_{j \in L_B}\gamma_j,
\end{split}
\end{equation}
where $K^{(12)} = -\kappa' t^{12}/E_c^{11}$ and $L_B$ is a closed loop shown in Fig.~\ref{MCB_scheme} and $P$ runs over each plaquette. $X_P$ for a green plaquette will be

\begin{equation}
\begin{split}
X_P  = &\sum_{i}
\left(\gamma_{r,2}(\mathbf{r}_i) \gamma_{b,1}(\mathbf{r}_i) \gamma_{g,1}(\mathbf{r}_i)\gamma_{p,1}(\mathbf{r}_i)\right)
\left(-\gamma_{r,2}(\mathbf{r}_i + \delta_1) \gamma_{b,1}(\mathbf{r}_i + \delta_1) \gamma_{g,1}(\mathbf{r}_i + \delta_1)\gamma_{p,1}(\mathbf{r}_i + \delta_1)\right)\\
& \times \left(\gamma_{r,2}(\mathbf{r}_i + \delta_1 - \delta_3)\gamma_{b,1}(\mathbf{r}_i + \delta_1 - \delta_3)\gamma_{g,1}(\mathbf{r}_i + \delta_1 - \delta_3)\gamma_{p,1}(\mathbf{r}_i + \delta_1 - \delta_3)\right)\\
& \times \left(- \gamma_{r,2}(\mathbf{r}_i + \delta_1 - \delta_3 + \delta_2)\gamma_{b,1}(\mathbf{r}_i + \delta_1 - \delta_3 + \delta_2) \gamma_{g,1}(\mathbf{r}_i + \delta_1 - \delta_3 + \delta_2) \gamma_{p,1}(\mathbf{r}_i + \delta_1 - \delta_3 + \delta_2)\right)\\
& \times \left(\gamma_{r,2}(\mathbf{r}_i + \delta_2 - \delta_3)\gamma_{b,1}(\mathbf{r}_i + \delta_2 - \delta_3)\gamma_{g,1}(\mathbf{r}_i + \delta_2 - \delta_3) \gamma_{p,1}(\mathbf{r}_i + \delta_2 - \delta_3)\right)\\
& \times \left( \gamma_{r,2}(\mathbf{r}_i + \delta_2)\gamma_{b,1}(\mathbf{r}_i + \delta_2)\gamma_{g,1}(\mathbf{r}_i + \delta_2) \gamma_{p,1}(\mathbf{r}_i + \delta_2)\right),
\end{split}
\end{equation} 
which is written in term of pauli operators around the plaquette $L_B$:

\begin{equation}
\begin{split}
& X_P = \prod_{j \in \hexagon}\sigma_j^x.
\end{split}
\end{equation}

The next nontrivial term in twelfth order for a green plaquette reads as follows:

\begin{equation}
\begin{split}
H_{eff,C}^{(12)}  = &-\kappa'  \frac{t^{12}}{E_c^{11}} \sum_{i}
\left(\gamma_{r,2}(\mathbf{r}_i) \gamma_{r,2}(\mathbf{r}_i + \delta_1)\right)
\left(\gamma_{b,1}(\mathbf{r}_i + \delta_1 - \delta_3) \gamma_{b,1}(\mathbf{r}_i + \delta_1)\right)
\left(\gamma_{g,2}(\mathbf{r}_i + \delta_1 - \delta_3) \gamma_{g,2}(\mathbf{r}_i + \delta_1)\right)\\
& \times \left(\gamma_{p,1}(\mathbf{r}_i + \delta_1 - \delta_3) \gamma_{p,1}(\mathbf{r}_i + \delta_1)\right)
\left(\gamma_{r,2}(\mathbf{r}_i + \delta_1 - \delta_3) \gamma_{r,2}(\mathbf{r}_i + \delta_1 - \delta_3 + \delta_2)\right)
\left(\gamma_{b,1}(\mathbf{r}_i + \delta_2 - \delta_3) \gamma_{b,1}(\mathbf{r}_i + \delta_1 - \delta_3 + \delta_2) \right)\\
& \times \left(\gamma_{g,2}(\mathbf{r}_i + \delta_2 - \delta_3) \gamma_{g,2}(\mathbf{r}_i + \delta_1 - \delta_3 + \delta_2) \right)
\left(\gamma_{p,1}(\mathbf{r}_i + \delta_2 - \delta_3) \gamma_{p,1}(\mathbf{r}_i + \delta_1 - \delta_3 + \delta_2) \right)
\left(\gamma_{r,1}(\mathbf{r}_i + \delta_2 - \delta_3) \gamma_{r,1}(\mathbf{r}_i + \delta_2)\right)\\
& \times \left(\gamma_{b,1}(\mathbf{r}_i) \gamma_{b,1}(\mathbf{r}_i + \delta_2)\right)
\left(\gamma_{g,2}(\mathbf{r}_i) \gamma_{g,2}(\mathbf{r}_i + \delta_2)\right)
\left(\gamma_{p,1}(\mathbf{r}_i) \gamma_{p,1}(\mathbf{r}_i + \delta_2)\right).
\end{split}
\end{equation} 

It is written as
\begin{equation}
\begin{split}
H_{eff}^{(12)} = K^{(12)}\sum_{P}Y_P,~~~ Y_P =  \prod_{j \in L_C}\gamma_j.
\end{split}
\end{equation}
where $L_C$ is a closed loop shown in Fig.~\ref{MCB_scheme}. If we rearrange $Y_P$ as 
\begin{equation}
\begin{split}
Y_P  = &\sum_{i}
\left(\gamma_{r,2}(\mathbf{r}_i) \gamma_{b,1}(\mathbf{r}_i) \gamma_{g,2}(\mathbf{r}_i) \gamma_{p,1}(\mathbf{r}_i) \right)
\left(\gamma_{r,2}(\mathbf{r}_i + \delta_1)\gamma_{b,1}(\mathbf{r}_i + \delta_1)\gamma_{g,2}(\mathbf{r}_i + \delta_1) \gamma_{p,1}(\mathbf{r}_i + \delta_1)\right)\\
& \times \left(-\gamma_{r,2}(\mathbf{r}_i + \delta_1 - \delta_3)\gamma_{b,1}(\mathbf{r}_i + \delta_1 - \delta_3) \gamma_{g,2}(\mathbf{r}_i + \delta_1 - \delta_3)\gamma_{p,1}(\mathbf{r}_i + \delta_1 - \delta_3) \right)\\
& \times \left(\gamma_{r,2}(\mathbf{r}_i + \delta_1 - \delta_3 + \delta_2)\gamma_{b,1}(\mathbf{r}_i + \delta_1 - \delta_3 + \delta_2) \gamma_{g,2}(\mathbf{r}_i + \delta_1 - \delta_3 + \delta_2) \gamma_{p,1}(\mathbf{r}_i + \delta_1 - \delta_3 + \delta_2)\right)\\
& \times \left(-\gamma_{r,2}(\mathbf{r}_i + \delta_2 - \delta_3)\gamma_{b,1}(\mathbf{r}_i + \delta_2 - \delta_3) \gamma_{g,2}(\mathbf{r}_i + \delta_2 - \delta_3)\gamma_{p,1}(\mathbf{r}_i + \delta_2 - \delta_3)\right)\\
& \times \left( \gamma_{r,2}(\mathbf{r}_i + \delta_2) \gamma_{b,1}(\mathbf{r}_i + \delta_2)\gamma_{g,2}(\mathbf{r}_i + \delta_2) \gamma_{p,1}(\mathbf{r}_i + \delta_2)\right).
\end{split}
\end{equation} 
in terms of pauli operators it becomes

\begin{equation}
\begin{split}
 Y_P =  \prod_{j \in L_C}\sigma_j^y.
\end{split}
\end{equation}

Thus, we obtain the effective Hamiltonian in eq.\eqref{CC_H}:
\begin{equation}
\begin{split}
H_{eff}=K^{(6)} \sum_{P} Z_{P}+ K^{(12)} \sum_{P} \left(X_{P}+Y_{P}\right).
\end{split}
\end{equation}

\end{widetext}


\begin{thebibliography}{40}%
\makeatletter
\providecommand \@ifxundefined [1]{%
 \@ifx{#1\undefined}
}%
\providecommand \@ifnum [1]{%
 \ifnum #1\expandafter \@firstoftwo
 \else \expandafter \@secondoftwo
 \fi
}%
\providecommand \@ifx [1]{%
 \ifx #1\expandafter \@firstoftwo
 \else \expandafter \@secondoftwo
 \fi
}%
\providecommand \natexlab [1]{#1}%
\providecommand \enquote  [1]{``#1''}%
\providecommand \bibnamefont  [1]{#1}%
\providecommand \bibfnamefont [1]{#1}%
\providecommand \citenamefont [1]{#1}%
\providecommand \href@noop [0]{\@secondoftwo}%
\providecommand \href [0]{\begingroup \@sanitize@url \@href}%
\providecommand \@href[1]{\@@startlink{#1}\@@href}%
\providecommand \@@href[1]{\endgroup#1\@@endlink}%
\providecommand \@sanitize@url [0]{\catcode `\\12\catcode `\$12\catcode
  `\&12\catcode `\#12\catcode `\^12\catcode `\_12\catcode `\%12\relax}%
\providecommand \@@startlink[1]{}%
\providecommand \@@endlink[0]{}%
\providecommand \url  [0]{\begingroup\@sanitize@url \@url }%
\providecommand \@url [1]{\endgroup\@href {#1}{\urlprefix }}%
\providecommand \urlprefix  [0]{URL }%
\providecommand \Eprint [0]{\href }%
\providecommand \doibase [0]{http://dx.doi.org/}%
\providecommand \selectlanguage [0]{\@gobble}%
\providecommand \bibinfo  [0]{\@secondoftwo}%
\providecommand \bibfield  [0]{\@secondoftwo}%
\providecommand \translation [1]{[#1]}%
\providecommand \BibitemOpen [0]{}%
\providecommand \bibitemStop [0]{}%
\providecommand \bibitemNoStop [0]{.\EOS\space}%
\providecommand \EOS [0]{\spacefactor3000\relax}%
\providecommand \BibitemShut  [1]{\csname bibitem#1\endcsname}%
\let\auto@bib@innerbib\@empty
\bibitem [{\citenamefont {Nayak}\ \emph {et~al.}(2008)\citenamefont {Nayak},
  \citenamefont {Simon}, \citenamefont {Stern}, \citenamefont {Freedman},\ and\
  \citenamefont {Das~Sarma}}]{DasSarma:RMP2008}%
  \BibitemOpen
  \bibfield  {author} {\bibinfo {author} {\bibfnamefont {C.}~\bibnamefont
  {Nayak}}, \bibinfo {author} {\bibfnamefont {S.~H.}\ \bibnamefont {Simon}},
  \bibinfo {author} {\bibfnamefont {A.}~\bibnamefont {Stern}}, \bibinfo
  {author} {\bibfnamefont {M.}~\bibnamefont {Freedman}}, \ and\ \bibinfo
  {author} {\bibfnamefont {S.}~\bibnamefont {Das~Sarma}},\ }\href {\doibase
  10.1103/RevModPhys.80.1083} {\bibfield  {journal} {\bibinfo  {journal} {Rev.
  Mod. Phys.}\ }\textbf {\bibinfo {volume} {80}},\ \bibinfo {pages} {1083}
  (\bibinfo {year} {2008})}\BibitemShut {NoStop}%
\bibitem [{\citenamefont {Stormer}\ \emph {et~al.}(1999)\citenamefont
  {Stormer}, \citenamefont {Tsui},\ and\ \citenamefont {Gossard}}]{FQHE:1999}%
  \BibitemOpen
  \bibfield  {author} {\bibinfo {author} {\bibfnamefont {H.~L.}\ \bibnamefont
  {Stormer}}, \bibinfo {author} {\bibfnamefont {D.~C.}\ \bibnamefont {Tsui}}, \
  and\ \bibinfo {author} {\bibfnamefont {A.~C.}\ \bibnamefont {Gossard}},\
  }\href {\doibase 10.1103/RevModPhys.71.S298} {\bibfield  {journal} {\bibinfo
  {journal} {Rev. Mod. Phys.}\ }\textbf {\bibinfo {volume} {71}},\ \bibinfo
  {pages} {S298} (\bibinfo {year} {1999})}\BibitemShut {NoStop}%
\bibitem [{\citenamefont {Kasahara}\ \emph {et~al.}(2018)\citenamefont
  {Kasahara}, \citenamefont {Ohnishi}, \citenamefont {Mizukami}, \citenamefont
  {Tanaka}, \citenamefont {Ma}, \citenamefont {Sugii}, \citenamefont {Kurita},
  \citenamefont {Tanaka}, \citenamefont {Nasu}, \citenamefont {Motome},
  \citenamefont {Shibauchi},\ and\ \citenamefont {Matsuda}}]{Matsuda:Nat2018}%
  \BibitemOpen
  \bibfield  {author} {\bibinfo {author} {\bibfnamefont {Y.}~\bibnamefont
  {Kasahara}}, \bibinfo {author} {\bibfnamefont {T.}~\bibnamefont {Ohnishi}},
  \bibinfo {author} {\bibfnamefont {Y.}~\bibnamefont {Mizukami}}, \bibinfo
  {author} {\bibfnamefont {O.}~\bibnamefont {Tanaka}}, \bibinfo {author}
  {\bibfnamefont {S.}~\bibnamefont {Ma}}, \bibinfo {author} {\bibfnamefont
  {K.}~\bibnamefont {Sugii}}, \bibinfo {author} {\bibfnamefont
  {N.}~\bibnamefont {Kurita}}, \bibinfo {author} {\bibfnamefont
  {H.}~\bibnamefont {Tanaka}}, \bibinfo {author} {\bibfnamefont
  {J.}~\bibnamefont {Nasu}}, \bibinfo {author} {\bibfnamefont {Y.}~\bibnamefont
  {Motome}}, \bibinfo {author} {\bibfnamefont {T.}~\bibnamefont {Shibauchi}}, \
  and\ \bibinfo {author} {\bibfnamefont {Y.}~\bibnamefont {Matsuda}},\ }\href
  {https://doi.org/10.1038/s41586-018-0274-0} {\bibfield  {journal} {\bibinfo
  {journal} {Nature}\ }\textbf {\bibinfo {volume} {559}},\ \bibinfo {pages}
  {227} (\bibinfo {year} {2018})}\BibitemShut {NoStop}%
\bibitem [{\citenamefont {Hsieh}\ \emph {et~al.}(2016)\citenamefont {Hsieh},
  \citenamefont {Ishizuka}, \citenamefont {Balents},\ and\ \citenamefont
  {Hughes}}]{Hsieh2016}%
  \BibitemOpen
  \bibfield  {author} {\bibinfo {author} {\bibfnamefont {T.~H.}\ \bibnamefont
  {Hsieh}}, \bibinfo {author} {\bibfnamefont {H.}~\bibnamefont {Ishizuka}},
  \bibinfo {author} {\bibfnamefont {L.}~\bibnamefont {Balents}}, \ and\
  \bibinfo {author} {\bibfnamefont {T.~L.}\ \bibnamefont {Hughes}},\ }\href
  {\doibase 10.1103/PhysRevLett.116.086802} {\bibfield  {journal} {\bibinfo
  {journal} {Phys. Rev. Lett.}\ }\textbf {\bibinfo {volume} {116}},\ \bibinfo
  {pages} {086802} (\bibinfo {year} {2016})}\BibitemShut {NoStop}%
\bibitem [{\citenamefont {Hsieh}\ \emph {et~al.}(2017)\citenamefont {Hsieh},
  \citenamefont {Lu},\ and\ \citenamefont {Ludwig}}]{Hsieh2017}%
  \BibitemOpen
  \bibfield  {author} {\bibinfo {author} {\bibfnamefont {T.~H.}\ \bibnamefont
  {Hsieh}}, \bibinfo {author} {\bibfnamefont {Y.-M.}\ \bibnamefont {Lu}}, \
  and\ \bibinfo {author} {\bibfnamefont {A.~W.~W.}\ \bibnamefont {Ludwig}},\
  }\href {\doibase 10.1126/sciadv.1700729} {\bibfield  {journal} {\bibinfo
  {journal} {Science Advances}\ }\textbf {\bibinfo {volume} {3}} (\bibinfo
  {year} {2017}),\ 10.1126/sciadv.1700729},\ \Eprint
  {http://arxiv.org/abs/https://advances.sciencemag.org/content/3/10/e1700729.full.pdf}
  {https://advances.sciencemag.org/content/3/10/e1700729.full.pdf} \BibitemShut
  {NoStop}%
\bibitem [{\citenamefont {Terhal}\ \emph {et~al.}(2012)\citenamefont {Terhal},
  \citenamefont {Hassler},\ and\ \citenamefont {DiVincenzo}}]{Terhal:PRL2012}%
  \BibitemOpen
  \bibfield  {author} {\bibinfo {author} {\bibfnamefont {B.~M.}\ \bibnamefont
  {Terhal}}, \bibinfo {author} {\bibfnamefont {F.}~\bibnamefont {Hassler}}, \
  and\ \bibinfo {author} {\bibfnamefont {D.~P.}\ \bibnamefont {DiVincenzo}},\
  }\href {\doibase 10.1103/PhysRevLett.108.260504} {\bibfield  {journal}
  {\bibinfo  {journal} {Phys. Rev. Lett.}\ }\textbf {\bibinfo {volume} {108}},\
  \bibinfo {pages} {260504} (\bibinfo {year} {2012})}\BibitemShut {NoStop}%
\bibitem [{\citenamefont {Kitaev}(2003)}]{KITAEV20032}%
  \BibitemOpen
  \bibfield  {author} {\bibinfo {author} {\bibfnamefont {A.}~\bibnamefont
  {Kitaev}},\ }\href {\doibase https://doi.org/10.1016/S0003-4916(02)00018-0}
  {\bibfield  {journal} {\bibinfo  {journal} {Annals of Physics}\ }\textbf
  {\bibinfo {volume} {303}},\ \bibinfo {pages} {2} (\bibinfo {year}
  {2003})}\BibitemShut {NoStop}%
\bibitem [{\citenamefont {Nussinov}\ \emph {et~al.}(2012)\citenamefont
  {Nussinov}, \citenamefont {Ortiz},\ and\ \citenamefont
  {Cobanera}}]{Nussinov:prb2018}%
  \BibitemOpen
  \bibfield  {author} {\bibinfo {author} {\bibfnamefont {Z.}~\bibnamefont
  {Nussinov}}, \bibinfo {author} {\bibfnamefont {G.}~\bibnamefont {Ortiz}}, \
  and\ \bibinfo {author} {\bibfnamefont {E.}~\bibnamefont {Cobanera}},\ }\href
  {\doibase 10.1103/PhysRevB.86.085415} {\bibfield  {journal} {\bibinfo
  {journal} {Phys. Rev. B}\ }\textbf {\bibinfo {volume} {86}},\ \bibinfo
  {pages} {085415} (\bibinfo {year} {2012})}\BibitemShut {NoStop}%
\bibitem [{\citenamefont {Bombin}\ and\ \citenamefont
  {Martin-Delgado}(2006)}]{Bombin2006}%
  \BibitemOpen
  \bibfield  {author} {\bibinfo {author} {\bibfnamefont {H.}~\bibnamefont
  {Bombin}}\ and\ \bibinfo {author} {\bibfnamefont {M.~A.}\ \bibnamefont
  {Martin-Delgado}},\ }\href {\doibase 10.1103/PhysRevLett.97.180501}
  {\bibfield  {journal} {\bibinfo  {journal} {Phys. Rev. Lett.}\ }\textbf
  {\bibinfo {volume} {97}},\ \bibinfo {pages} {180501} (\bibinfo {year}
  {2006})}\BibitemShut {NoStop}%
\bibitem [{\citenamefont {Bombin}\ \emph {et~al.}(2009)\citenamefont {Bombin},
  \citenamefont {Kargarian},\ and\ \citenamefont
  {Martin-Delgado}}]{Bombin2009}%
  \BibitemOpen
  \bibfield  {author} {\bibinfo {author} {\bibfnamefont {H.}~\bibnamefont
  {Bombin}}, \bibinfo {author} {\bibfnamefont {M.}~\bibnamefont {Kargarian}}, \
  and\ \bibinfo {author} {\bibfnamefont {M.~A.}\ \bibnamefont
  {Martin-Delgado}},\ }\href {\doibase 10.1103/PhysRevB.80.075111} {\bibfield
  {journal} {\bibinfo  {journal} {Phys. Rev. B}\ }\textbf {\bibinfo {volume}
  {80}},\ \bibinfo {pages} {075111} (\bibinfo {year} {2009})}\BibitemShut
  {NoStop}%
\bibitem [{\citenamefont {Kargarian}\ \emph {et~al.}(2010)\citenamefont
  {Kargarian}, \citenamefont {Bombin},\ and\ \citenamefont
  {Martin-Delgado}}]{Kargarian_2010}%
  \BibitemOpen
  \bibfield  {author} {\bibinfo {author} {\bibfnamefont {M.}~\bibnamefont
  {Kargarian}}, \bibinfo {author} {\bibfnamefont {H.}~\bibnamefont {Bombin}}, \
  and\ \bibinfo {author} {\bibfnamefont {M.~A.}\ \bibnamefont
  {Martin-Delgado}},\ }\href {\doibase 10.1088/1367-2630/12/2/025018}
  {\bibfield  {journal} {\bibinfo  {journal} {New Journal of Physics}\ }\textbf
  {\bibinfo {volume} {12}},\ \bibinfo {pages} {025018} (\bibinfo {year}
  {2010})}\BibitemShut {NoStop}%
\bibitem [{\citenamefont {Kitaev}(2006)}]{KITAEV20062}%
  \BibitemOpen
  \bibfield  {author} {\bibinfo {author} {\bibfnamefont {A.}~\bibnamefont
  {Kitaev}},\ }\href {\doibase https://doi.org/10.1016/j.aop.2005.10.005}
  {\bibfield  {journal} {\bibinfo  {journal} {Annals of Physics}\ }\textbf
  {\bibinfo {volume} {321}},\ \bibinfo {pages} {2} (\bibinfo {year} {2006})},\
  \bibinfo {note} {january Special Issue}\BibitemShut {NoStop}%
\bibitem [{\citenamefont {Qi}\ \emph {et~al.}(2009)\citenamefont {Qi},
  \citenamefont {Hughes}, \citenamefont {Raghu},\ and\ \citenamefont
  {Zhang}}]{Qi2009}%
  \BibitemOpen
  \bibfield  {author} {\bibinfo {author} {\bibfnamefont {X.-L.}\ \bibnamefont
  {Qi}}, \bibinfo {author} {\bibfnamefont {T.~L.}\ \bibnamefont {Hughes}},
  \bibinfo {author} {\bibfnamefont {S.}~\bibnamefont {Raghu}}, \ and\ \bibinfo
  {author} {\bibfnamefont {S.-C.}\ \bibnamefont {Zhang}},\ }\href {\doibase
  10.1103/PhysRevLett.102.187001} {\bibfield  {journal} {\bibinfo  {journal}
  {Phys. Rev. Lett.}\ }\textbf {\bibinfo {volume} {102}},\ \bibinfo {pages}
  {187001} (\bibinfo {year} {2009})}\BibitemShut {NoStop}%
\bibitem [{\citenamefont {Zhang}\ \emph {et~al.}(2013)\citenamefont {Zhang},
  \citenamefont {Kane},\ and\ \citenamefont {Mele}}]{Zhang2013}%
  \BibitemOpen
  \bibfield  {author} {\bibinfo {author} {\bibfnamefont {F.}~\bibnamefont
  {Zhang}}, \bibinfo {author} {\bibfnamefont {C.~L.}\ \bibnamefont {Kane}}, \
  and\ \bibinfo {author} {\bibfnamefont {E.~J.}\ \bibnamefont {Mele}},\ }\href
  {\doibase 10.1103/PhysRevLett.111.056402} {\bibfield  {journal} {\bibinfo
  {journal} {Phys. Rev. Lett.}\ }\textbf {\bibinfo {volume} {111}},\ \bibinfo
  {pages} {056402} (\bibinfo {year} {2013})}\BibitemShut {NoStop}%
\bibitem [{\citenamefont {B\'eri}\ and\ \citenamefont
  {Cooper}(2012)}]{Beri2012}%
  \BibitemOpen
  \bibfield  {author} {\bibinfo {author} {\bibfnamefont {B.}~\bibnamefont
  {B\'eri}}\ and\ \bibinfo {author} {\bibfnamefont {N.~R.}\ \bibnamefont
  {Cooper}},\ }\href {\doibase 10.1103/PhysRevLett.109.156803} {\bibfield
  {journal} {\bibinfo  {journal} {Phys. Rev. Lett.}\ }\textbf {\bibinfo
  {volume} {109}},\ \bibinfo {pages} {156803} (\bibinfo {year}
  {2012})}\BibitemShut {NoStop}%
\bibitem [{\citenamefont {Altland}\ and\ \citenamefont
  {Egger}(2013)}]{Altland2013}%
  \BibitemOpen
  \bibfield  {author} {\bibinfo {author} {\bibfnamefont {A.}~\bibnamefont
  {Altland}}\ and\ \bibinfo {author} {\bibfnamefont {R.}~\bibnamefont
  {Egger}},\ }\href {\doibase 10.1103/PhysRevLett.110.196401} {\bibfield
  {journal} {\bibinfo  {journal} {Phys. Rev. Lett.}\ }\textbf {\bibinfo
  {volume} {110}},\ \bibinfo {pages} {196401} (\bibinfo {year}
  {2013})}\BibitemShut {NoStop}%
\bibitem [{\citenamefont {B\'eri}(2013)}]{Beri2013}%
  \BibitemOpen
  \bibfield  {author} {\bibinfo {author} {\bibfnamefont {B.}~\bibnamefont
  {B\'eri}},\ }\href {\doibase 10.1103/PhysRevLett.110.216803} {\bibfield
  {journal} {\bibinfo  {journal} {Phys. Rev. Lett.}\ }\textbf {\bibinfo
  {volume} {110}},\ \bibinfo {pages} {216803} (\bibinfo {year}
  {2013})}\BibitemShut {NoStop}%
\bibitem [{\citenamefont {Zazunov}\ \emph {et~al.}(2014)\citenamefont
  {Zazunov}, \citenamefont {Altland},\ and\ \citenamefont
  {Egger}}]{Zazunov2014}%
  \BibitemOpen
  \bibfield  {author} {\bibinfo {author} {\bibfnamefont {A.}~\bibnamefont
  {Zazunov}}, \bibinfo {author} {\bibfnamefont {A.}~\bibnamefont {Altland}}, \
  and\ \bibinfo {author} {\bibfnamefont {R.}~\bibnamefont {Egger}},\ }\href
  {\doibase 10.1088/1367-2630/16/1/015010} {\bibfield  {journal} {\bibinfo
  {journal} {New J. Phys.}\ }\textbf {\bibinfo {volume} {16}},\ \bibinfo
  {pages} {015010} (\bibinfo {year} {2014})}\BibitemShut {NoStop}%
\bibitem [{\citenamefont {Altland}\ \emph {et~al.}(2014)\citenamefont
  {Altland}, \citenamefont {B\'eri}, \citenamefont {Egger},\ and\ \citenamefont
  {Tsvelik}}]{Altland2014}%
  \BibitemOpen
  \bibfield  {author} {\bibinfo {author} {\bibfnamefont {A.}~\bibnamefont
  {Altland}}, \bibinfo {author} {\bibfnamefont {B.}~\bibnamefont {B\'eri}},
  \bibinfo {author} {\bibfnamefont {R.}~\bibnamefont {Egger}}, \ and\ \bibinfo
  {author} {\bibfnamefont {A.~M.}\ \bibnamefont {Tsvelik}},\ }\href {\doibase
  10.1103/PhysRevLett.113.076401} {\bibfield  {journal} {\bibinfo  {journal}
  {Phys. Rev. Lett.}\ }\textbf {\bibinfo {volume} {113}},\ \bibinfo {pages}
  {076401} (\bibinfo {year} {2014})}\BibitemShut {NoStop}%
\bibitem [{\citenamefont {Lutchyn}\ \emph {et~al.}(2010)\citenamefont
  {Lutchyn}, \citenamefont {Sau},\ and\ \citenamefont
  {Das~Sarma}}]{Lutchyn2010}%
  \BibitemOpen
  \bibfield  {author} {\bibinfo {author} {\bibfnamefont {R.~M.}\ \bibnamefont
  {Lutchyn}}, \bibinfo {author} {\bibfnamefont {J.~D.}\ \bibnamefont {Sau}}, \
  and\ \bibinfo {author} {\bibfnamefont {S.}~\bibnamefont {Das~Sarma}},\ }\href
  {\doibase 10.1103/PhysRevLett.105.077001} {\bibfield  {journal} {\bibinfo
  {journal} {Phys. Rev. Lett.}\ }\textbf {\bibinfo {volume} {105}},\ \bibinfo
  {pages} {077001} (\bibinfo {year} {2010})}\BibitemShut {NoStop}%
\bibitem [{\citenamefont {Alicea}(2012)}]{Alicea2012}%
  \BibitemOpen
  \bibfield  {author} {\bibinfo {author} {\bibfnamefont {J.}~\bibnamefont
  {Alicea}},\ }\href@noop {} {\bibfield  {journal} {\bibinfo  {journal} {Rep.
  Prog. Phys.}\ } (\bibinfo {year} {2012})}\BibitemShut {NoStop}%
\bibitem [{\citenamefont {Fu}(2010)}]{Fu2010}%
  \BibitemOpen
  \bibfield  {author} {\bibinfo {author} {\bibfnamefont {L.}~\bibnamefont
  {Fu}},\ }\href {\doibase 10.1103/PhysRevLett.104.056402} {\bibfield
  {journal} {\bibinfo  {journal} {Phys. Rev. Lett.}\ }\textbf {\bibinfo
  {volume} {104}},\ \bibinfo {pages} {056402} (\bibinfo {year}
  {2010})}\BibitemShut {NoStop}%
\bibitem [{\citenamefont {Zazunov}\ \emph {et~al.}(2011)\citenamefont
  {Zazunov}, \citenamefont {Yeyati},\ and\ \citenamefont
  {Egger}}]{Zazunov2011}%
  \BibitemOpen
  \bibfield  {author} {\bibinfo {author} {\bibfnamefont {A.}~\bibnamefont
  {Zazunov}}, \bibinfo {author} {\bibfnamefont {A.~L.}\ \bibnamefont {Yeyati}},
  \ and\ \bibinfo {author} {\bibfnamefont {R.}~\bibnamefont {Egger}},\ }\href
  {\doibase 10.1103/PhysRevB.84.165440} {\bibfield  {journal} {\bibinfo
  {journal} {Phys. Rev. B}\ }\textbf {\bibinfo {volume} {84}},\ \bibinfo
  {pages} {165440} (\bibinfo {year} {2011})}\BibitemShut {NoStop}%
\bibitem [{\citenamefont {Landau}\ \emph {et~al.}(2016)\citenamefont {Landau},
  \citenamefont {Plugge}, \citenamefont {Sela}, \citenamefont {Altland},
  \citenamefont {Albrecht},\ and\ \citenamefont {Egger}}]{Plugge2016}%
  \BibitemOpen
  \bibfield  {author} {\bibinfo {author} {\bibfnamefont {L.~A.}\ \bibnamefont
  {Landau}}, \bibinfo {author} {\bibfnamefont {S.}~\bibnamefont {Plugge}},
  \bibinfo {author} {\bibfnamefont {E.}~\bibnamefont {Sela}}, \bibinfo {author}
  {\bibfnamefont {A.}~\bibnamefont {Altland}}, \bibinfo {author} {\bibfnamefont
  {S.~M.}\ \bibnamefont {Albrecht}}, \ and\ \bibinfo {author} {\bibfnamefont
  {R.}~\bibnamefont {Egger}},\ }\href {\doibase 10.1103/PhysRevLett.116.050501}
  {\bibfield  {journal} {\bibinfo  {journal} {Phys. Rev. Lett.}\ }\textbf
  {\bibinfo {volume} {116}},\ \bibinfo {pages} {050501} (\bibinfo {year}
  {2016})}\BibitemShut {NoStop}%
\bibitem [{\citenamefont {Plugge}\ \emph
  {et~al.}(2016{\natexlab{a}})\citenamefont {Plugge}, \citenamefont {Landau},
  \citenamefont {Sela}, \citenamefont {Altland}, \citenamefont {Flensberg},\
  and\ \citenamefont {Egger}}]{Plugge2016_2}%
  \BibitemOpen
  \bibfield  {author} {\bibinfo {author} {\bibfnamefont {S.}~\bibnamefont
  {Plugge}}, \bibinfo {author} {\bibfnamefont {L.~A.}\ \bibnamefont {Landau}},
  \bibinfo {author} {\bibfnamefont {E.}~\bibnamefont {Sela}}, \bibinfo {author}
  {\bibfnamefont {A.}~\bibnamefont {Altland}}, \bibinfo {author} {\bibfnamefont
  {K.}~\bibnamefont {Flensberg}}, \ and\ \bibinfo {author} {\bibfnamefont
  {R.}~\bibnamefont {Egger}},\ }\href {\doibase 10.1103/PhysRevB.94.174514}
  {\bibfield  {journal} {\bibinfo  {journal} {Phys. Rev. B}\ }\textbf {\bibinfo
  {volume} {94}},\ \bibinfo {pages} {174514} (\bibinfo {year}
  {2016}{\natexlab{a}})}\BibitemShut {NoStop}%
\bibitem [{\citenamefont {Plugge}\ \emph
  {et~al.}(2016{\natexlab{b}})\citenamefont {Plugge}, \citenamefont {Zazunov},
  \citenamefont {Eriksson}, \citenamefont {Tsvelik},\ and\ \citenamefont
  {Egger}}]{Plugge2016_3}%
  \BibitemOpen
  \bibfield  {author} {\bibinfo {author} {\bibfnamefont {S.}~\bibnamefont
  {Plugge}}, \bibinfo {author} {\bibfnamefont {A.}~\bibnamefont {Zazunov}},
  \bibinfo {author} {\bibfnamefont {E.}~\bibnamefont {Eriksson}}, \bibinfo
  {author} {\bibfnamefont {A.~M.}\ \bibnamefont {Tsvelik}}, \ and\ \bibinfo
  {author} {\bibfnamefont {R.}~\bibnamefont {Egger}},\ }\href {\doibase
  10.1103/PhysRevB.93.104524} {\bibfield  {journal} {\bibinfo  {journal} {Phys.
  Rev. B}\ }\textbf {\bibinfo {volume} {93}},\ \bibinfo {pages} {104524}
  (\bibinfo {year} {2016}{\natexlab{b}})}\BibitemShut {NoStop}%
\bibitem [{\citenamefont {Thomson}\ and\ \citenamefont
  {Pientka}(2018)}]{thomson2018}%
  \BibitemOpen
  \bibfield  {author} {\bibinfo {author} {\bibfnamefont {A.}~\bibnamefont
  {Thomson}}\ and\ \bibinfo {author} {\bibfnamefont {F.}~\bibnamefont
  {Pientka}},\ }\href@noop {} {\enquote {\bibinfo {title} {Simulating spin
  systems with majorana networks},}\ } (\bibinfo {year} {2018}),\ \Eprint
  {http://arxiv.org/abs/1807.09291} {arXiv:1807.09291 [cond-mat.mes-hall]}
  \BibitemShut {NoStop}%
\bibitem [{\citenamefont {Tsvelik}(2020)}]{Tsvelik2020}%
  \BibitemOpen
  \bibfield  {author} {\bibinfo {author} {\bibfnamefont {A.}~\bibnamefont
  {Tsvelik}},\ }\href {\doibase 10.1103/physrevlett.125.197202} {\bibfield
  {journal} {\bibinfo  {journal} {Physical Review Letters}\ }\textbf {\bibinfo
  {volume} {125}} (\bibinfo {year} {2020}),\
  10.1103/physrevlett.125.197202}\BibitemShut {NoStop}%
\bibitem [{\citenamefont {Bravyi}\ \emph {et~al.}(2010)\citenamefont {Bravyi},
  \citenamefont {Terhal},\ and\ \citenamefont {Leemhuis}}]{Bravyi2010}%
  \BibitemOpen
  \bibfield  {author} {\bibinfo {author} {\bibfnamefont {S.}~\bibnamefont
  {Bravyi}}, \bibinfo {author} {\bibfnamefont {B.~M.}\ \bibnamefont {Terhal}},
  \ and\ \bibinfo {author} {\bibfnamefont {B.}~\bibnamefont {Leemhuis}},\
  }\href {\doibase 10.1088/1367-2630/12/8/083039} {\bibfield  {journal}
  {\bibinfo  {journal} {New Journal of Physics}\ }\textbf {\bibinfo {volume}
  {12}},\ \bibinfo {pages} {083039} (\bibinfo {year} {2010})}\BibitemShut
  {NoStop}%
\bibitem [{\citenamefont {Kargarian}(2008)}]{Kargarian2008}%
  \BibitemOpen
  \bibfield  {author} {\bibinfo {author} {\bibfnamefont {M.}~\bibnamefont
  {Kargarian}},\ }\href {\doibase 10.1103/PhysRevA.78.062312} {\bibfield
  {journal} {\bibinfo  {journal} {Phys. Rev. A}\ }\textbf {\bibinfo {volume}
  {78}},\ \bibinfo {pages} {062312} (\bibinfo {year} {2008})}\BibitemShut
  {NoStop}%
\bibitem [{\citenamefont {Schnyder}\ \emph {et~al.}(2008)\citenamefont
  {Schnyder}, \citenamefont {Ryu}, \citenamefont {Furusaki},\ and\
  \citenamefont {Ludwig}}]{Schnyder2008}%
  \BibitemOpen
  \bibfield  {author} {\bibinfo {author} {\bibfnamefont {A.~P.}\ \bibnamefont
  {Schnyder}}, \bibinfo {author} {\bibfnamefont {S.}~\bibnamefont {Ryu}},
  \bibinfo {author} {\bibfnamefont {A.}~\bibnamefont {Furusaki}}, \ and\
  \bibinfo {author} {\bibfnamefont {A.~W.~W.}\ \bibnamefont {Ludwig}},\ }\href
  {\doibase 10.1103/PhysRevB.78.195125} {\bibfield  {journal} {\bibinfo
  {journal} {Phys. Rev. B}\ }\textbf {\bibinfo {volume} {78}},\ \bibinfo
  {pages} {195125} (\bibinfo {year} {2008})}\BibitemShut {NoStop}%
\bibitem [{\citenamefont {Camjayi}\ \emph {et~al.}(2017)\citenamefont
  {Camjayi}, \citenamefont {Arrachea}, \citenamefont {Aligia},\ and\
  \citenamefont {von Oppen}}]{Camjayi2017}%
  \BibitemOpen
  \bibfield  {author} {\bibinfo {author} {\bibfnamefont {A.}~\bibnamefont
  {Camjayi}}, \bibinfo {author} {\bibfnamefont {L.}~\bibnamefont {Arrachea}},
  \bibinfo {author} {\bibfnamefont {A.}~\bibnamefont {Aligia}}, \ and\ \bibinfo
  {author} {\bibfnamefont {F.}~\bibnamefont {von Oppen}},\ }\href {\doibase
  10.1103/PhysRevLett.119.046801} {\bibfield  {journal} {\bibinfo  {journal}
  {Phys. Rev. Lett.}\ }\textbf {\bibinfo {volume} {119}},\ \bibinfo {pages}
  {046801} (\bibinfo {year} {2017})}\BibitemShut {NoStop}%
\bibitem [{\citenamefont {Nakosai}\ \emph {et~al.}(2012)\citenamefont
  {Nakosai}, \citenamefont {Tanaka},\ and\ \citenamefont
  {Nagaosa}}]{Nakosai2012}%
  \BibitemOpen
  \bibfield  {author} {\bibinfo {author} {\bibfnamefont {S.}~\bibnamefont
  {Nakosai}}, \bibinfo {author} {\bibfnamefont {Y.}~\bibnamefont {Tanaka}}, \
  and\ \bibinfo {author} {\bibfnamefont {N.}~\bibnamefont {Nagaosa}},\ }\href
  {\doibase 10.1103/PhysRevLett.108.147003} {\bibfield  {journal} {\bibinfo
  {journal} {Phys. Rev. Lett.}\ }\textbf {\bibinfo {volume} {108}},\ \bibinfo
  {pages} {147003} (\bibinfo {year} {2012})}\BibitemShut {NoStop}%
\bibitem [{\citenamefont {Deng}\ \emph {et~al.}(2012)\citenamefont {Deng},
  \citenamefont {Viola},\ and\ \citenamefont {Ortiz}}]{Deng2012}%
  \BibitemOpen
  \bibfield  {author} {\bibinfo {author} {\bibfnamefont {S.}~\bibnamefont
  {Deng}}, \bibinfo {author} {\bibfnamefont {L.}~\bibnamefont {Viola}}, \ and\
  \bibinfo {author} {\bibfnamefont {G.}~\bibnamefont {Ortiz}},\ }\href
  {\doibase 10.1103/PhysRevLett.108.036803} {\bibfield  {journal} {\bibinfo
  {journal} {Phys. Rev. Lett.}\ }\textbf {\bibinfo {volume} {108}},\ \bibinfo
  {pages} {036803} (\bibinfo {year} {2012})}\BibitemShut {NoStop}%
\bibitem [{\citenamefont {Wong}\ and\ \citenamefont {Law}(2012)}]{Wong2012}%
  \BibitemOpen
  \bibfield  {author} {\bibinfo {author} {\bibfnamefont {C.~L.~M.}\
  \bibnamefont {Wong}}\ and\ \bibinfo {author} {\bibfnamefont {K.~T.}\
  \bibnamefont {Law}},\ }\href {\doibase 10.1103/PhysRevB.86.184516} {\bibfield
   {journal} {\bibinfo  {journal} {Phys. Rev. B}\ }\textbf {\bibinfo {volume}
  {86}},\ \bibinfo {pages} {184516} (\bibinfo {year} {2012})}\BibitemShut
  {NoStop}%
\bibitem [{\citenamefont {Keselman}\ \emph {et~al.}(2013)\citenamefont
  {Keselman}, \citenamefont {Fu}, \citenamefont {Stern},\ and\ \citenamefont
  {Berg}}]{Keselman2013}%
  \BibitemOpen
  \bibfield  {author} {\bibinfo {author} {\bibfnamefont {A.}~\bibnamefont
  {Keselman}}, \bibinfo {author} {\bibfnamefont {L.}~\bibnamefont {Fu}},
  \bibinfo {author} {\bibfnamefont {A.}~\bibnamefont {Stern}}, \ and\ \bibinfo
  {author} {\bibfnamefont {E.}~\bibnamefont {Berg}},\ }\href {\doibase
  10.1103/PhysRevLett.111.116402} {\bibfield  {journal} {\bibinfo  {journal}
  {Phys. Rev. Lett.}\ }\textbf {\bibinfo {volume} {111}},\ \bibinfo {pages}
  {116402} (\bibinfo {year} {2013})}\BibitemShut {NoStop}%
\bibitem [{\citenamefont {Wen}(2003)}]{Wen:PRL2003}%
  \BibitemOpen
  \bibfield  {author} {\bibinfo {author} {\bibfnamefont {X.-G.}\ \bibnamefont
  {Wen}},\ }\href {\doibase 10.1103/PhysRevLett.90.016803} {\bibfield
  {journal} {\bibinfo  {journal} {Phys. Rev. Lett.}\ }\textbf {\bibinfo
  {volume} {90}},\ \bibinfo {pages} {016803} (\bibinfo {year}
  {2003})}\BibitemShut {NoStop}%
\bibitem [{\citenamefont {Scheurer}\ and\ \citenamefont
  {Schmalian}(2015)}]{Scheurer2015}%
  \BibitemOpen
  \bibfield  {author} {\bibinfo {author} {\bibfnamefont {M.~S.}\ \bibnamefont
  {Scheurer}}\ and\ \bibinfo {author} {\bibfnamefont {J.}~\bibnamefont
  {Schmalian}},\ }\href {\doibase 10.1038/ncomms7005} {\bibfield  {journal}
  {\bibinfo  {journal} {Nature Communications}\ }\textbf {\bibinfo {volume}
  {6}},\ \bibinfo {pages} {6005} (\bibinfo {year} {2015})}\BibitemShut
  {NoStop}%
\bibitem [{\citenamefont {Sato}\ \emph {et~al.}(2008)\citenamefont {Sato},
  \citenamefont {Souma}, \citenamefont {Nakayama}, \citenamefont {Terashima},
  \citenamefont {Sugawara}, \citenamefont {Takahashi}, \citenamefont
  {Kamihara}, \citenamefont {Hirano},\ and\ \citenamefont {Hosono}}]{Sato2008}%
  \BibitemOpen
  \bibfield  {author} {\bibinfo {author} {\bibfnamefont {T.}~\bibnamefont
  {Sato}}, \bibinfo {author} {\bibfnamefont {S.}~\bibnamefont {Souma}},
  \bibinfo {author} {\bibfnamefont {K.}~\bibnamefont {Nakayama}}, \bibinfo
  {author} {\bibfnamefont {K.}~\bibnamefont {Terashima}}, \bibinfo {author}
  {\bibfnamefont {K.}~\bibnamefont {Sugawara}}, \bibinfo {author}
  {\bibfnamefont {T.}~\bibnamefont {Takahashi}}, \bibinfo {author}
  {\bibfnamefont {Y.}~\bibnamefont {Kamihara}}, \bibinfo {author}
  {\bibfnamefont {M.}~\bibnamefont {Hirano}}, \ and\ \bibinfo {author}
  {\bibfnamefont {H.}~\bibnamefont {Hosono}},\ }\href {\doibase
  10.1143/jpsj.77.063708} {\bibfield  {journal} {\bibinfo  {journal} {Journal
  of the Physical Society of Japan}\ }\textbf {\bibinfo {volume} {77}},\
  \bibinfo {pages} {063708} (\bibinfo {year} {2008})}\BibitemShut {NoStop}%
\bibitem [{\citenamefont {Lutchyn}\ \emph {et~al.}(2018)\citenamefont
  {Lutchyn}, \citenamefont {Bakkers}, \citenamefont {Kouwenhoven},
  \citenamefont {Krogstrup}, \citenamefont {Marcus},\ and\ \citenamefont
  {Oreg}}]{Lutchyn:NatRev2018}%
  \BibitemOpen
  \bibfield  {author} {\bibinfo {author} {\bibfnamefont {R.~M.}\ \bibnamefont
  {Lutchyn}}, \bibinfo {author} {\bibfnamefont {E.~P. A.~M.}\ \bibnamefont
  {Bakkers}}, \bibinfo {author} {\bibfnamefont {L.~P.}\ \bibnamefont
  {Kouwenhoven}}, \bibinfo {author} {\bibfnamefont {P.}~\bibnamefont
  {Krogstrup}}, \bibinfo {author} {\bibfnamefont {C.~M.}\ \bibnamefont
  {Marcus}}, \ and\ \bibinfo {author} {\bibfnamefont {Y.}~\bibnamefont
  {Oreg}},\ }\href {https://doi.org/10.1038/s41578-018-0003-1} {\bibfield
  {journal} {\bibinfo  {journal} {Nature Reviews Materials}\ }\textbf {\bibinfo
  {volume} {3}},\ \bibinfo {pages} {52} (\bibinfo {year} {2018})}\BibitemShut
  {NoStop}%
\end{thebibliography}
%

\end{document}